\documentclass[12pt]{article}
\usepackage{a4}
\usepackage{amsmath,amssymb}
\usepackage[dvips]{graphicx}
\usepackage{epsfig}
\usepackage{amsfonts}

\makeatletter

\@addtoreset{equation}{section}
\makeatother

\newcommand{\tr}{\text{tr}\, }
\newcommand{\vev}[1]{\left\langle #1 \right\rangle}
\newcommand{\nn}{\nonumber}

\begin{document}
\begin{titlepage}
\rightline{\today}

\vspace{30mm}

\begin{center}
{\Large A Renormalization Group Approach \\
 to  A Yang-Mills Two Matrix Model}  \\

\vspace{20mm}
Shoichi Kawamoto$^1$ and Dan Tomino$^2$
	\\
	{\footnotesize\it Department of Physics, Tunghai University,
Taichung 40704, Taiwan}\\
{\footnotesize\tt ${}^1$kawamoto@thu.edu.tw,
${}^2$dantomino@thu.edu.tw}

\vspace{10mm}
{\bf Abstract}
\end{center}
A Yang-Mills type two matrix model with
mass terms is studied by use of
a matrix renormalization group approach proposed by Brezin and Zinn-Justin.
The renormalization group method indicates that
the model exhibits a critical behavior
similar to that of two dimensional Euclidean gravity.
A massless limit and
the generation of quadratic terms 
along the renormalization group flow are discussed.     
\end{titlepage}

\newpage
\allowdisplaybreaks

\section{Introduction}

A couple of decades ago, Brezin and Zinn-Justin proposed a renormalization
group (RG) approach to matrix models
in the large-$N$ limit \cite{Brezin:1992yc}.
With a given action for an $N \times N$ matrix,
this matrix RG flow is defined through the integration over a part of
$N\times N$ matrix degrees of freedom which results in an $(N-1) \times (N-1)$
matrix with a deformed action. 
The change of the action may be interpreted as a flow along the change
of the matrix size, $N \rightarrow N-1$.
In the large-$N$ limit, this flow is approximately described by a
differential equation of Callan-Synamzik type,
with beta functions for the changes of
the coupling constants under the procedure.
For a simple class of hermitian one-matrix models,
it is found that, near a fixed point of these beta functions, 
a solution of the flow equation approximates the critical behaviors
which were analytically calculated in
\cite{Brezin:1977sv, Kazakov:1989bc}.
For example, in \cite{Brezin:1992yc}, the $\phi^4$ matrix model and 
hermitian one-matrix models with generic even polynomial potentials are discussed
based on lower order calculations in perturbation theory.
The results are qualitatively reasonable but quantitatively
not so accurate compared to the known exact result.
Later in \cite{Higuchi:1994rv}, the reparametrization invariance
of the model is dealt with by use of the Schwinger-Dyson equations,
and their results show fine agreement with the 
exact results quantitatively.
The great advantage of the matrix RG approach is its applicability to a wide
class of matrix models which are not exactly solvable.
Especially, as described so far, this method enables us to, at least
qualitatively, understand the critical behavior of various classes of
matrix models in the large-$N$ limit.

The critical behaviors of hermitian one and two matrix models
are used to define two-dimensional Euclidean quantum gravity coupled
to conformal matters \cite{David:1984tx}
and $D < 2$ non-critical string theories \cite{Douglas:1989ve}.
Therefore, we expect to explore the critical behavior of
wider classes of matrix models, such as a $q$-state Potts model, and,
thus, understand the nature of two dimensional quantum gravity
on a more general ground\footnote{
For the application to $c=1$ matrix models, see, for example, \cite{Dasgupta:2003kk}.}.
\\

In this paper, 
we apply the matrix RG approach to a Yang-Mills (YM) type two-matrix model with mass terms
\begin{eqnarray}
S=\tr
\left( \frac{m}{2}\hat{A}^2+\frac{M}{2}\hat{B}^2-\frac{g}{2}[\hat{A},\hat{B}]^2  \right),
\label{YMmm}
\end{eqnarray}
where $\hat{A}$ and $\hat{B}$ are $N\times N$ hermitian matrices.
There are several motivations to consider this model.
First of all, this model belongs to a class of the model which we cannot solve
exactly by use of conventional approach to the matrix models\footnote{%
As discussed in the later section,
this model, with only two matrices,  can be analytically investigated 
by use of several specialized techniques \cite{Kazakov:1998ji}.
However, these analytic answers are not so convenient to analyze,
while the application of the matrix RG to this model
is as simple as  that to one matrix models, as we will see.
The eigenvalue distribution of this model
has been recently studied in \cite{Filev:2013pza}, 
which gives a consistent result with \cite{Kazakov:1998ji}.}.
By using the matrix RG approach, 
we may extract some qualitative features of critical behavior in the
YM-type matrix models.
As long as we employ perturbation theory to derive the RG
equation, it is also easy to apply to various YM-type multi-matrix models.

Another motivation to investigate this model
comes from a massless limit $m,M \rightarrow 0$
in (\ref{YMmm}).
In this limit, the YM matrix model is being
a large-$N$ reduced model of two dimensional pure Yang-Mills theory\footnote{%
The two dimensional reduced model is not well-defined unless the mass terms are
introduced. We come back to this point when we consider the massless limit.}
and can be regarded as a bosonic part of
the D-instanton action or IKKT matrix model \cite{Ishibashi:1996xs}.
Especially, the latter is a promising candidate of the constructive definition of
superstring theory.
We expect the matrix RG approach to be useful to probe the large-$N$ dynamics
of the IKKT matrix model and D-branes,
and then non-perturbative dynamics of superstring theory.  \\

This paper is organized as follows.
First, we review the matrix RG approach in Section
\ref{sec:matr-renorm-group},
by taking the $\phi^4$ one-matrix model as a simple example.
In Section \ref{sec:phi4-model},
we carry out the matrix RG analysis for this $\phi^4$ model.
We start with perturbative calculation to some lower orders,
which leads to reasonable but numerically not so accurate results,
and then discuss a couple of trials for improvement,
such as including higher order corrections,
taking account of reparametrization invariance,
and the existence of nontrivial saddle points.
In \cite{Higuchi:1994rv},
some of these improvements have first been introduced and the
results are successful 
for one and two matrix models.
However, there, the authors take advantage of solvable nature of these
models, which is not available in general YM matrix models.
We cultivate methods for improvement which are applicable more
generally.
After these preparations, we apply the matrix RG to the YM type two-matrix
model \eqref{YMmm} in Section \ref{sec:yang-mills-matrix}.
We again start with perturbation theory
and then consider improvements.
The result suggests that the model exhibits a similar critical
behavior to the one-matrix model.
In \cite{Kazakov:1998ji}, the model \eqref{YMmm} was dissected
analytically and its critical behavior is also discussed.
We discuss the relation to our results.
In Section \ref{sec:massless-limit},
we consider a massless limit and connection to matrix RG of IKKT-type matrix models.
Section \ref{sec:summary-discussion} is devoted to summary and
discussion.
Several appendices are served for supplemental explanations. \\

Before closing the introduction, we remark that 
the most of the analysis in Section \ref{sec:yang-mills-matrix}
can straightforwardly be carried out for
a more general model
\begin{eqnarray}
S=\tr\left( \frac{m}{2}\hat{A}^2+\frac{M}{2}\hat{B}^2
+g\hat{A}^2\hat{B}^2+h\hat{A}\hat{B}\hat{A} \hat{B}
+ \frac{p}{4}\hat{A}^4+\frac{q}{4}\hat{B}^4  \right) .
\label{Ymmm2}
\end{eqnarray}
We occasionally refer to this model.
Note that this model coincides with \eqref{YMmm}
when $p=q=0$ and $h=-g$.


\section{A Matrix Renormalization Group Equation}
\label{sec:matr-renorm-group}
In this section, we briefly summarize the procedure
of the matrix RG. Our treatment is close to \cite{Higuchi:1994rv}. 
For simplicity, we take the $\phi^4$ matrix model
\begin{eqnarray}
S= \tr \left( \frac{1}{2}\hat{\phi}^2+\frac{g}{4}\hat{\phi}^4 \right)
\label{p4action}
\end{eqnarray}
as an example. 
Here $\hat\phi$ is an $N \times N$ hermitian matrix.
The partition function and the free energy are defined by
\begin{eqnarray}
Z_N(g) = \int D\hat{\phi}\;e^{-N S[\hat\phi]} \,,
\label{pf}
\\
F(N,g)=- \frac{1}{N^2}\log Z_N(g) \,.
\label{fegergy}
\end{eqnarray}

In the matrix RG, we integrate out a part of matrix degrees of
freedom. This is an analogue of the coarse-graining in
the usual Wilsonian RG procedure.
We decompose $\hat{\phi}$ as
\begin{eqnarray*}
\hat{\phi}=\left(\begin{array}{cc}
\phi & v \\
v^{\dagger} & \alpha
\end{array}\right), \qquad
v_i=\hat{\phi}_{iN}, \quad v_i^{\dagger}=\hat{\phi}_{Ni},\quad  \alpha=\hat{\phi}_{NN},
\end{eqnarray*}
where $\phi$ is an $(N-1)\times (N-1)$ hermitian matrix,
$v$ an $(N-1)$-vector, and $\alpha$ a real number.
Under this decomposition, the partition function (\ref{pf}) is 
\begin{eqnarray}
&&
Z_N=\int D\phi dv dv^{\dagger}d\alpha
\;\exp\left[
-N\left(
S[\phi]
+\frac{1}{2}\alpha^2+\frac{g}{4}\alpha^4
+v^{\dagger}Yv+\frac{g}{2}|v|^2|v|^2 
\right) \right] \label{pf2},
 \nonumber 
\end{eqnarray}  
where $Y={\bf 1}+g(\alpha^2{\bf 1}+\alpha\phi+\phi^2)$
and $|v|^2 = \delta^{ij} v_i v^{\dagger}_j$.
With integrating out $v$, $v^{\dagger}$, and $\alpha$,  we define
\begin{align}
  e^{-NV[\phi]} =&
\int dv dv^{\dagger}d\alpha
\;\exp\left[
-N\left(
\frac{1}{2}\alpha^2+\frac{g}{4}\alpha^4
+v^{\dagger}Yv+\frac{g}{2}|v|^2|v|^2 
\right) \right]
\,.
\label{eq:5}
\end{align}
Therefore
the matrix analogue of the coarse-graining procedure
results in the change of the action,
$S[\hat{\phi}]\rightarrow S[\phi]+V[\phi]$.
In the perturbation theory of
$v$ and $\alpha$, $V[\phi]$ is obtained as a sum of the connected vacuum
diagrams.
In general, $V[\phi]$ involves infinity many terms including various
multi-trace terms.
To derive the RG equation, we consider 
\begin{eqnarray}
\frac{ Z_{N} }{ Z_{N-1}}= \left\langle  e^{-S[\phi]-NV[\phi]}\right\rangle,
\label{toRG}
\end{eqnarray}
where $\langle \cdots \rangle =Z_{N-1}^{-1} \int D\phi (\cdots) e^{-(N-1)S[\phi]}$,
namely, the expectation value with respect to the smaller size matrix
of the original action.
We will use the same notation in the case of the YM-type matrix models.
In the large-$N$ limit,
we use the large $N$ factorization property for the gauge invariant
operators\footnote{%
Now ``$\tr$''  stands for the trace of $N-1\times N-1$ matrices.
We use the same symbol but it would not cause any confusion.},
\begin{eqnarray}
 \vev{ \frac{1}{N}\tr{\cal O}_1  \frac{1}{N}\tr{\cal O}_2 }
= \vev{ \frac{1}{N}\tr{\cal O}_1}  \vev{ \frac{1}{N}\tr{\cal O}_2} +{\cal O}(N^{-2})
\end{eqnarray}  
and the right hand side of (\ref{toRG}) 
becomes
$\vev{e^{-S[\phi]-NV[\phi]}} = e^{-\vev{S[\phi]}-N \vev{V[\phi]}}$ at
the leading order.
Using (\ref{fegergy}), the logarithm of (\ref{toRG}) leads to an RG equation
\begin{eqnarray}
\left(N\frac{\partial}{\partial N}+2 \right)F(N,g)=\vev{V[\phi]} \,.
\end{eqnarray}
We have replaced $F(N,g)-F(N-1,g)$ as the differentiation with respect
to $N$ and have dropped subleading terms in the $1/N$ expansion.
In general, $V$ has the form, 
\begin{eqnarray}
V[\phi] = v_0(g)+v_2(g) \tr \phi^2 + v_4(g)\tr\phi^4+ v_\text{higher}(g,\phi).
\end{eqnarray}
As we will see later,
 $v_2 \tr \phi^2$ term can be absorbed into $(N-1)S[\phi]$ by a
suitable rescaling of $\phi$ to make the quadratic term
canonical, $\frac{N-1}{2}\tr \phi^2$.
$v_\text{higher}(g, \phi)$ in general
involves infinitely many higher order terms as well as multi-trace terms
that did not exist in the original action (\ref{p4action}),
and it introduces non-linear terms in the RG equation.
In the main part of this paper,
we frequently consider the linear RG equation with
$v_\text{higher}(g,\phi)$ dropped.
Without $v_\text{higher}(g,\phi)$ term, the linearized RG equation reads
\begin{eqnarray}   
\left(N\frac{\partial}{\partial N}+2\right)F(N,g)=
r(g)+\beta (g)\frac{\partial F}{\partial g}.
\label{RGeq}
\end{eqnarray}
Recall that $\vev{\frac{1}{N}\tr\phi^4}=4\frac{\partial F}{\partial g}$.
The beta function $\beta(g)$ is described by $v_4(g)$ together with the rescaling
factor just mentioned, and is interpreted as
\begin{eqnarray}
\beta(g)=N\frac{\partial g'}{\partial N} \,,
\end{eqnarray}
where $g'$ is the ``new'' coupling constant for $\tr \phi^4$ term
appearing after the integration of $v$, $v^\dagger$ and $\alpha$.

A fixed point of the RG equation is given by a zero of the beta function,
$\beta(g_c)=0$.
Near a fixed point $g\sim g_c$,
the linearized RG equation (\ref{RGeq}) determines  a non-analytic behavior of the free energy,
\begin{eqnarray}
F(g,N)\sim (g-g_c)^{\gamma}f\left( (g-g_c)N^{\frac{2}{\gamma}} \right) \,, 
\label{Fsing}
\end{eqnarray}  
where $f$ is a undetermined function.
The singularity is characterized by the exponent $\gamma$ given by
\begin{eqnarray}
\gamma=\frac{2}{\beta'(g_c)},
\end{eqnarray}  
where $\beta'(g)=\frac{\partial \beta}{\partial g}$. 
This singular behavior is identical to the one for the $c<1$ matrix model
which describes 2D gravity coupled to conformal matters.
Through this correspondence, $\gamma$ is identified with
the critical exponent of the $c<1$ model,
\begin{eqnarray}
\gamma=\frac{1}{12}\left[25-c+\sqrt{(1-c)(25-c)}\right].
\end{eqnarray} 

If there are non-linear terms in the RG equation (\ref{RGeq}),
one can employ the procedure presented in \cite{Higuchi:1994rv} to calculate the exponent.
In Appendix \ref{sec:non-linear},
we briefly summarize this method and apply to the first nonlinear term
we deal with in this paper.

\section{The $\phi^4$ Matrix Model}
\label{sec:phi4-model}

In this section, we study the matrix RG of the $\phi^4$ matrix model. 
We use the decomposition in (\ref{pf2}).
Integration of $v$, $v^{\dagger}$, and $\alpha$ is carried out by using perturbation theory,
with the propagators 
\begin{eqnarray}
\langle v_iv^{\dagger}_j\rangle_0 = \frac{1}{N}Y^{-1}_{ij}=
\frac{1}{N}\left[ \frac{1}{{\bf 1}+g(\alpha^2{\bf 1}+\alpha\phi+\phi^2) }\right]_{ij}, 
\qquad
\langle \alpha\alpha \rangle_0 =\frac{1}{N}, 
\end{eqnarray}
and the interaction vertices
\begin{eqnarray}
\frac{gN}{2}|v|^2|v|^2, \quad \frac{gN}{4}\alpha^4.
\end{eqnarray}

The solution of the $\phi^4$ model is known to be given by an
eigenvalue distribution with a single branch cut and
to exhibit a critical behavior
with the exponent $\gamma=\frac{5}{2}$,
which corresponds to the $c=0$ (pure) 2D gravity.
With the normalization of the action (\ref{p4action}),
the critical coupling constant is given by $g_c=-\frac{1}{12}$. 
We will test our approximation by comparing it with these exact results.

\subsection{Lower order calculations}

First, we carry out a standard perturbative calculation to obtain $V[\phi]$.
We will drop perturbative contributions from $\langle \alpha\alpha\rangle_0$
which are subleading in the large-$N$ limit.
The integral with respect to $\alpha$ may have a nontrivial saddle point
that would take an effect to modify $V[\phi]$.
We postpone discussion about the effects from an induced  $\alpha$-potential,
and first consider the perturbation around the trivial vacuum $\alpha=0$.
To the second order in $g$, we obtain
\begin{eqnarray}
N S[\hat{\phi}] \, \rightarrow \,
N \tr \left(\frac{1}{2}\phi^2+\frac{g}{4}\phi^4\right)
+NP_0(g)+P_2(g)\, \tr\phi^2-gP_4(g)\, \tr\phi^4,
\label{s+v}
\end{eqnarray} 
where 
\begin{eqnarray}
P_0(g)=\frac{1}{2}(g-g^2), \quad P_{2}(g)=g-g^2, \quad P_4(g)=\frac{g}{2}.
\label{lowcorre}
\end{eqnarray}
Subleading contributions in large $N$ limit are discarded. 
Next, we rescale the matrix $\phi$ so that the quadratic term has
the canonical form as an $(N-1) \times (N-1)$ matrix,
\begin{eqnarray}
\frac{N}{2}\left(1+\frac{2P_2}{N}\right)\tr \phi^2 \rightarrow
\frac{N-1}{2} \tr \phi^2
\,.
\label{resc}
\end{eqnarray}   
After the rescaling, (\ref{s+v}) becomes, 
\begin{eqnarray}
N \left(P_0+P_2+\frac{1}{2}\right)
+(N-1) \tr\left[\frac{1}{2}\phi^2+\frac{g}{4}\left(1-\frac{1+4P_2+4P_4}{N}\right)\phi^4\right]
+\mathcal{O}(1) \,.
\end{eqnarray}
$P_2+\frac{1}{2}$ in the first parenthesis comes from the change of the measure $D\phi$
due to the rescaling.
Now the coefficient of the $\frac{1}{4}\tr \phi^4$ term is understood
as a ``new'' coupling
constant modified by the coarse-graining procedure,
and  we obtain
\begin{eqnarray}
r(g)&=& P_0+P_2+\frac{1}{2}=\frac{3}{2}(g-g^2)+\frac{1}{2}, \\
\beta(g)&=&-g(1+4P_2+4P_4)=-g(1+6g-4g^2).
\end{eqnarray} 
Therefore, a fixed point of the beta function and the corresponding
critical exponent are evaluated as
\begin{eqnarray}
g_c= \frac{3-\sqrt{13}}{4} \simeq -0.151,  \qquad
\gamma=\frac{4\sqrt{13}}{13(-3+\sqrt{13})}\simeq 1.83
\,,
\end{eqnarray}
where we have chosen the root with the smallest non-vanishing absolute value
as $g_c$.
From the viewpoint of perturbation theory, it would correspond to the first
singularity to be realized\footnote{%
For a general class of matrix models, in the large-$N$ limit,
various kinds of eigenvalue distribution and accordingly various
types of critical behavior may be realized.
Therefore, in general,
it is not obvious that which fixed point of the matrix RG for a given model
corresponds to which critical behavior of the matrix model in question.
In the case of the $\phi^4$ model,
we naturally expect that the matrix RG recovers
the singularity produced by the standard one-cut solution.}.
Comparing them with the exact values $g_c=-\frac{1}{12}$ and
$\gamma=\frac{5}{2}$ calculated in \cite{Brezin:1977sv},
we find that this lower order calculation provides a reasonable,
but quantitatively not so satisfactory, result.

\subsection{Higher order corrections}

Next, we take the effect of higher order corrections into account and study 
how it changes the lower order result.
We also trim the higher order terms by use of reparametrization
invariance and observe improvement.

\paragraph{The exact propagator in the planar limit:}

Instead of pushing perturbative calculation to higher orders,
we introduce the exact (or full) propagator of $v$ in the large-$N$
limit.
The exact propagator contains all the self-energy corrections,
and then the vacuum diagrams in terms of this exact propagator gets
simplified.
Actually, to the leading order of the large-$N$ limit,
it turns our that the simplification is drastic and it is possible
to make a partial resummation.

Let $\frac{1}{N}C_{ij}$ be the exact propagator of $v_i$ and
$v^{\dagger}_j$
\begin{align}
  \label{eq:1}
  \frac{1}{N}C_{ij} = \vev{v_i v_j^\dagger}
= \,
  \raisebox{-2pt}{\includegraphics[width=5em]{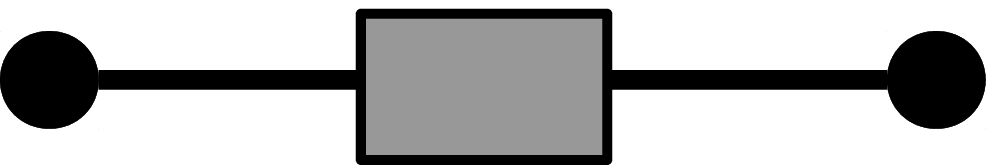}} \,,
\end{align}
where the box represents the exact propagator
and the expectation value involves the full integral over $v$ and $v^\dagger$.
 The exact propagator can be described by following consistency
 equation,
\begin{eqnarray}
\frac{C}{N}
= \,
  \raisebox{-2.5pt}{\includegraphics[width=6em]{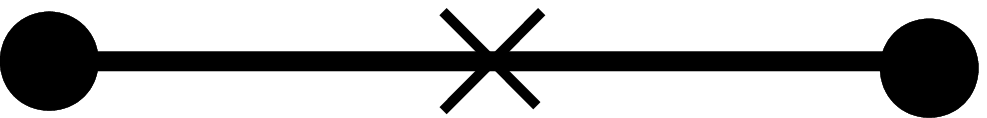}} \,
+ \, \raisebox{-6.8pt}{\includegraphics[width=7.5em]{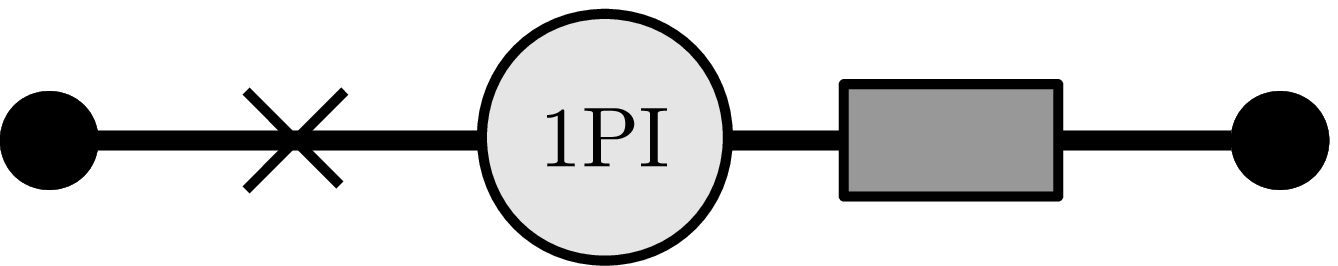}} \,
=\frac{Y^{-1}}{N}+\frac{Y^{-1}}{N}D^{1PI}\frac{C}{N}
\,,
\label{psd}
\end{eqnarray} 
where $Y^{-1}/N$ (a crossed line) is the tree level propagator
and $D^{1PI}$ represents the amputated two-point one-particle
irreducible (1PI) diagrams.
To the leading order in the large-$N$ limit, 
there is only one diagram in terms of the exact propagator
 and
it has a particularly simple expression,
\begin{align}
  \label{eq:2}
  \raisebox{-11pt}{\includegraphics[width=6em]{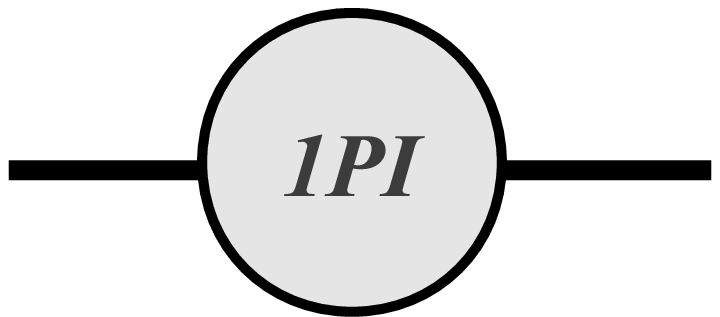}} \, =&
\, \raisebox{-1pt}{\includegraphics[width=7em]{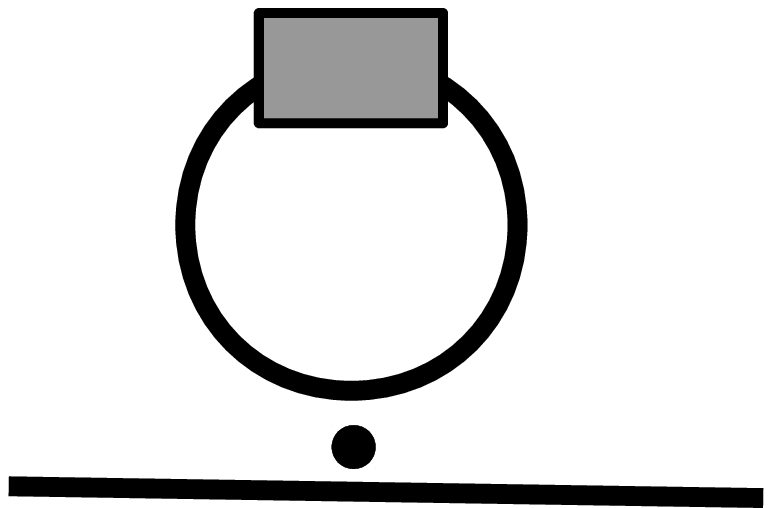}} 
\quad  \rightarrow \quad   D^{1PI}=-g {\tr C} +\mathcal{O}(1) . 
\end{align}
Here the interaction vertex
$\frac{gN}{2}|v|^2|v|^2=\frac{gN}{2}(v_iv^\dagger_i)(v_j v^\dagger_j)$ is represented by
the four lines with a black dot, to make the flow of the
indices $i$ and $j$ manifest.
Therefore, we have the self-consistency equation,
\begin{eqnarray}
C=Y^{-1}-gY^{-1}\frac{\tr C}{N} C \,.
\label{sdp2}
\end{eqnarray}
The exact $v$-integral is obtained by considering the
following planar connected vacuum diagrams in terms of the exact
propagator in the large-$N$ limit,
\begin{align}
  \label{eq:3}
   \raisebox{-8pt}{\includegraphics[width=2em]{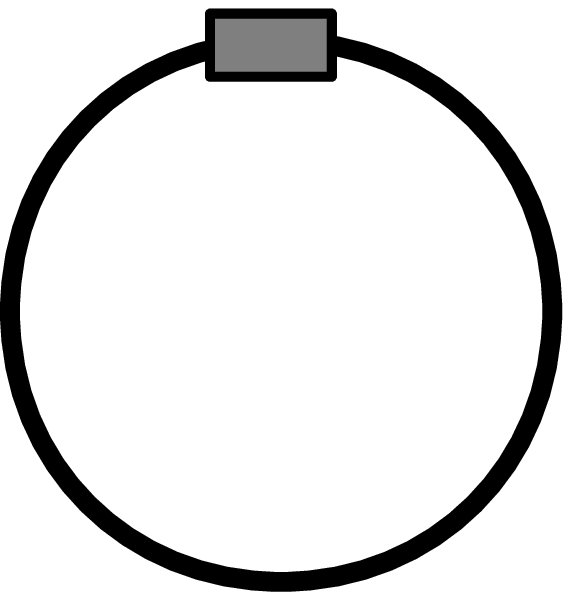}}
  \, - \,
 \raisebox{15pt}{\includegraphics[clip, angle=-90, width=5em]{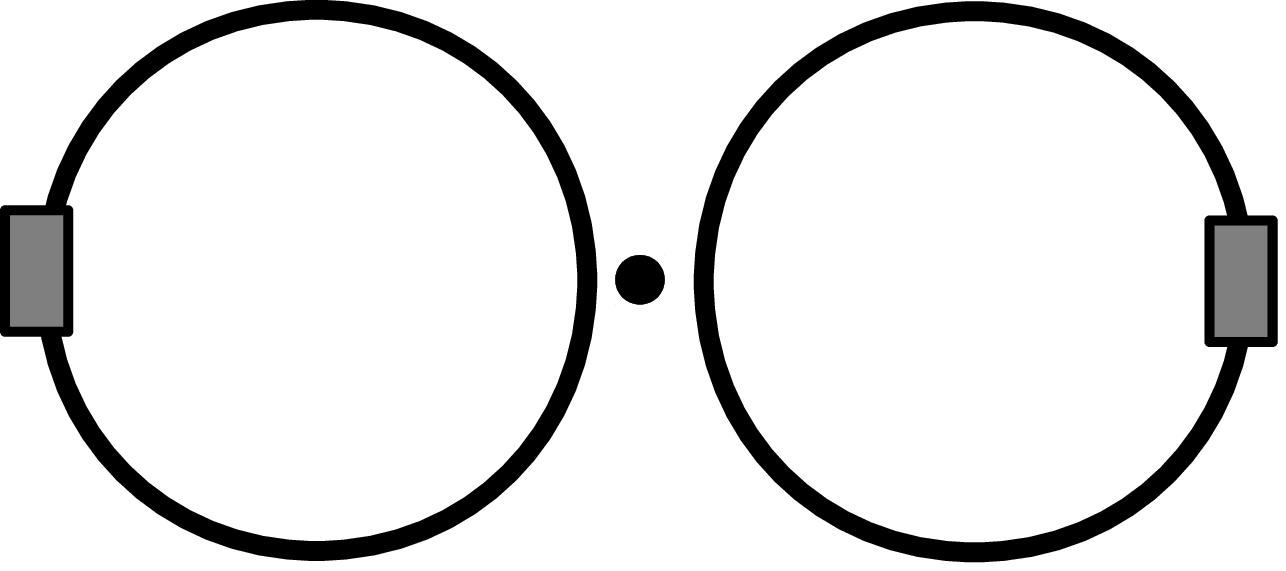}} \,
 =&
 -\frac{1}{N} \tr \ln {C} - \frac{g}{2} \left(\frac{1}{N} \tr C \right)^2
+ \mathcal{O}(N^{-1}) \,.
\end{align}
By solving (\ref{sdp2}) iteratively and plugging it into
(\ref{eq:3}), one can check that (\ref{eq:3}) reproduces the
perturbative result, and it confirms the negative sign for the second term.

Unfortunately, we have not been able to obtain a full solution of (\ref{sdp2}).
However, if we neglect the contribution of $\phi$, namely setting
$\phi=0$, the tree level propagator gets simplified as
$Y^{-1}=\frac{1}{1+g \alpha^2} \mathbf{1}$.
In this case, we can assume $C=c \mathbf{1}$ and solve (\ref{sdp2}) as
\begin{align}
  c=& \frac{1+g\alpha^2}{2g} \left(
-1 + \sqrt{1+\frac{4g}{(1+g\alpha^2)^2}}
\right) \,,
\label{solc}
\end{align}
where the sign is chosen so that $g=0$ case comes back to the tree
level result.
This is the resummed propagator that does not include the self-energy corrections
from the interaction vertices 
$gN(\alpha v^{\dagger}\phi v)$ and $gN(v^{\dagger}\phi^2v)$
but include those from the self-interaction as well as $gN( v^\dagger \alpha^2 v)$ interaction.
One can use this $c$ to improve (\ref{s+v}). 
Let us calculate the change of the action to the third order of
perturbation theory by use of $c$,
\begin{align}
N S[\hat{\phi}] \rightarrow &
N \tr\left(\frac{1}{2}\phi^2+\frac{g}{4}\phi^4\right)
+NP_0(g)+P_2(g)\tr\phi^2-gP_4(g)\tr\phi^4
\nn\\ &
+g^2P_6(g)\tr\phi^6 
+\frac{P_{2,2}(g)}{N}\tr\phi^2 \tr\phi^2,
\label{s+v2}
\end{align}
where\footnote{%
Here the vacuum energy part $P_0$ is obtained by substituting
$C=c\mathbf{1}$ into (\ref{eq:3}). 
The negative sign for $\mathcal{O}(g)$ term may also be understood in the
following way: we can introduce a subsidiary variable $h$ to rewrite $v^4$
vertex as $\frac{-gN}{2}(h^2-2v^{\dagger}hv)$, at least formally.
Integrating out $v$, with $\alpha$ and $\phi$ neglected, leads
an effective action $-\frac{gN}{2}h^2+N\log (1+gh)$.
Evaluating this by its saddle point value, $h=(1+gh)^{-1}$,
one finds $-N \log h -\frac{Ng}{2}h^2$. Finally, recalling that
the saddle point value of $h$ is $h=|v|^2$ at the beginning,
namely $h$ gives the exact propagator that does not
involve $\phi$ nor $\alpha$, one can confirm that $P_0$ indeed is
 the vacuum energy part.}
\begin{eqnarray}
\hspace{-3mm}
P_0(g)=-\log c-\frac{gc^2}{2},
 \quad P_2(g)=gc, \quad P_4(g)=\frac{gc^2}{2}, \quad
 P_6(g)=\frac{gc^3}{3}, \quad P_{2,2}(g)=\frac{g^3c^4}{2},
\end{eqnarray}
and  $c=\frac{1}{2g}(-1+\sqrt{1+4g})$.

To compare this resummed result with the one in the previous subsection, 
we take only the first line of (\ref{s+v2}), namely up to 
the second order of perturbation.
After rescaling $\phi$ as before, we obtain  
\begin{eqnarray}
\beta(g)=-g\left( 1+4gc+2gc^2 \right) \,.
\end{eqnarray} 
The non-zero fixed point of the beta function and the critical exponent of it
are evaluated numerically as 
\begin{eqnarray}
g_c \simeq -0.132, 
\qquad
\gamma=\frac{-2\sqrt{1+4g_c}}{\sqrt{1+4g_c}+12g_c} \simeq 1.52 .
\end{eqnarray} 
By employing the partially-exact propagator $c$, 
we achieve some improvement, but not so significant one.

\paragraph{Schwinger-Dyson equations and higher order terms:}

So far, we have ignored higher order terms induced by the
coarse-graining procedure. 
As pointed out in \cite{Higuchi:1994rv, Higuchi:1993pu},
such higher order terms are not independent of one another,
especially related to the terms in the original action,
due to the large reparametrization invariance of the matrix model.
Therefore, simply neglecting higher order terms
would not be a good approximation.
We now introduce the Schwinger-Dyson equations from the reparametrization invariance
and trim the higher order terms, by following \cite{Higuchi:1994rv}.

The reparametrization invariance implies that the correlation functions are invariant under
the redefinition of the matrix, $\phi \rightarrow \phi + \epsilon \phi^n$.
The invariance of the partition function leads to the following Schwinger-Dyson equations,
\begin{align}
  \label{eq:4}
\int D\phi\; \tr\left[ \frac{\delta}{\delta \phi}
  \left(\phi^n e^{-NS} \right)\right] =0 \,.
\end{align}

To deal with $\tr \phi^6$ and $(\tr \phi^2)^2$ terms in (\ref{s+v2}),
we use the following Schwinger-Dyson equations that are from $n=1$ and $n=3$ ones
respectively,
\begin{align}
\vev{1} =&  \vev{\frac{\tr\phi^2}{N}} + g\vev{ \frac{\tr\phi^4}{N}} \,,
\\
2\vev{ \frac{\tr\phi^2}{N}}
+\vev{ \frac{\tr\phi}{N} \; \frac{\tr\phi}{N}}
 =& \vev{ \frac{\tr\phi^4}{N}} + g\vev{\frac{\tr\phi^6}{N}} \,.
\end{align}
Thanks to the large-$N$ factorization property, 
$N^{-2} \vev{\tr\phi \; \tr\phi } = N^{-1} \vev{ \tr\phi}\, N^{-1} \vev{ \tr\phi} +\mathcal{O}(N^{-1})
\sim 0$ to the leading order of the large-$N$ limit, because
 $\vev{\tr \phi}=0 $.
Therefore $\tr \phi^6$ and $(\tr \phi^2)^2$ terms can be identified with
\begin{align}
\vev{ \tr \phi^6} =& \frac{2}{g}\vev{ \tr\phi^2 } -\frac{1}{g}\vev{ \tr\phi^4}, \\
\vev{ \tr\phi^2 }^2 =& \left( N-g \vev{\tr  \phi^4 } \right)^2. 
\label{eq:6}
\end{align}
Due to an appearance of the $\vev{\tr \phi^4 }^2$ term, we obtain an RG
equation of the free energy with a non-linear term
\begin{eqnarray}
\left(N\frac{\partial }{\partial N}+2\right)F=
b_0(g)+b_1(g)\frac{\partial F}{\partial g}+b_{2}(g)\left(\frac{\partial F}{\partial g}\right)^2,
\label{nlRG}
\end{eqnarray}
where
\begin{eqnarray}
b_0(g)&=&\frac{1}{2}+P_0+P_{2}+2gP_6+P_{2,2}, \quad \nonumber \\
b_1(g)&=&-g\Big\{1+4(P_2+P_4+P_6+2gP_6+2P_{2,2} ) \Big\}, \quad \nonumber \\
b_{2}(g)&=&16g^2P_{2,2}.
\label{nlbetas}
\end{eqnarray}
Note that all $P_{2}$, $P_4$ and $P_6$ start with $O(g)$ terms.
In fact, all $P_{n}$ from $n$-th order perturbation calculations
would contribute to $b_1$. 
At the linearized level (namely the $b_2$ part dropped),
a fixed point of the beta function and the corresponding exponent
are calculated as
\begin{eqnarray}
g_c\simeq -0.113, \qquad \gamma\simeq 1.59.
\end{eqnarray}
In Appendix \ref{sec:non-linear},
we calculate these quantities with the $b_2$ term by use of the method
developed in \cite{Higuchi:1994rv},
and obtain $\gamma_\text{nonlinear}\simeq 1.59$.
The non-linear term in (\ref{nlRG}) dose not change the result from the linear level analysis.
Since $b_2$ is proportional to $g^5$,
it seems natural that the change is not visible for this small value of $g_c$.   
\\

As we have seen, 
some improvement has been achieved to include
higher order terms in perturbation theory as well as
to take into account of reparametrization invariance,
and we
find that the value of the fixed point by the matrix RG approaches to
the exact value $\frac{-1}{12}$ within $35.1$\% relative error.
On the other hand,
the approximated values the critical exponent $\gamma$ take the
closest number to the exact one $\gamma=5/2$ when we simply use a lower
order calculation.
To this order, the improvement through the exact propagator
and the Schwinger-Dyson equations does not appear to be
so significant.

As we go further higher orders in perturbation theory,
more complicated terms are to be induced,
and we expect further improvement.
At the sixth order, a triple trace term $\tr \phi^2 \tr\phi^2
\tr\phi^2$ appears, and
it gives a further nonlinear $\left(\frac{\partial F}{\partial g}\right)^3$ term in the
RG equation
through the Schwinger-Dyson equations and the large-$N$ factorization.

\subsection{The effect of nontrivial saddle points of $\alpha$}

So far, we have considered the effective action for $\phi$
around an obvious saddle point of $\alpha$ potential, $\alpha=0$, in \eqref{eq:5}.
Although perturbative contribution from the $\alpha$ propagator 
$\langle \alpha\alpha\rangle$ is subleading in the large-$N$ limit,
therefore negligible,
other saddle points of the $\alpha$-potential would have significant effect,
which we will investigate in the following.
For simplicity, we restrict ourselves in calculation up to $\mathcal{O}(g^2)$.

In \eqref{eq:5}, one can carry out only $v$-part integration
which leads an effective action for $\phi$ and $\alpha$,
\begin{eqnarray}
V[\phi,\alpha]=V[\phi,0]
+ \frac{1}{2}\left(
1+2L_2-2gK\frac{\tr\phi^2}{N} \right) 
\alpha^2+\frac{g}{4}(1-4L_4)\alpha^4,
\label{apotential}
\end{eqnarray}     
where $V[\phi,0]$ denotes an $\alpha$ independent part which
coincides with $V[\phi]$ that we have considered in the previous subsections, and
\begin{eqnarray}
L_2=g-g^2, \quad L_4=\frac{g}{2}, \quad K=\frac{3g}{2}.
\end{eqnarray}
It should be noted that we have dropped single trace terms with odd powers in $\phi$,
like $\tr \phi^{2n+1}$, as they will vanish after we take the expectation value with respect to
$S[\phi]$.

Now the question is whether there is a saddle point $\alpha_c$ in
the effective action \eqref{apotential} so that $V[\phi,\alpha_c] <
V[\phi,0]$.
If so, the contribution of perturbation theory around
$\alpha=\alpha_c$ dominates
the path-integral as $\exp [-N(V[\phi,0]-V[\phi,\alpha_c])]$ is
exponentially small in the large-$N$ limit.
In (\ref{apotential}), such a non-trivial saddle point appears when
the coefficient of the quadratic term of $\alpha$ is negative,
and simultaneously the coefficient for the quartic term positive.
The latter condition implies $0<g<1/2$.
To estimate the signature of the quadratic term coefficient, we employ
perturbation theory with respect to $S[\phi]$, namely use 
$\vev{\frac{\tr\phi^2}{N}}=1-2g+9g^2+\mathcal{O}(g^3)$. 
To the second order of perturbation in $g$,
it turns out that the evaluated coefficient becomes negative for
$g<-0.290$ or $0.690<g$.
Therefore, to this order, there does not appear a
non-trivial saddle point for the $\alpha$-effective potential.

Although the relevance of non-trivial saddle points is not
justified at this order, we continue observing the effect of the
non-trivial saddle point, to present how it
affects the result if it existed.
We may also assume an optimistic attitude as it would become significant at
higher orders\footnote{%
Since perturbative corrections
  appear with $(-g)^n$,  a next order correction to $L_4$ has the negative
  sign. 
It thus leaves a possibility that, for some positive or negative values of $g$, there may
appear a non-trivial saddle point}.
The saddle point value of the effective action is then
\begin{eqnarray}
V[\phi,\alpha_c]= 
V[\phi,0] -\frac{\left(1+2L_2-2gK\frac{\tr\phi^2}{N}\right)^2}{4g(1-4L_4)}.
\end{eqnarray}
The fluctuation of $\alpha$ around this vacuum is negligible as before.
It produces new $\tr\phi^2$ and $\tr\phi^2\tr\phi^2$ terms
that change the beta functions.
Including them to (\ref{s+v}),
with the Schwinger-Dyson equation \eqref{eq:4},
 leads to an RG-equation 
\begin{eqnarray}
\left(N\frac{\partial}{\partial N}+2\right)F=b_0+b_1\frac{\partial F}{\partial g}+b_2\left(\frac{\partial F}{\partial g}\right)^2,
\end{eqnarray}
with
\begin{eqnarray}
b_0&=& \frac{1}{2} + P_0+P_2-\frac{1}{1-4L_4} \left(
\frac{(1+2L_2)^2}{4g}-K(1+2L_2)+gK^2 \right),
\\ 
b_1&=&-g\left(1+4P_2+4P_4+4K\frac{1+2L_2}{1-4L_4}-\frac{8gK^2}{1-4L_4}\right), \\
b_2&=&-\frac{16g^3K^2}{1-4L_4},
\end{eqnarray}
where $P_0$, $P_2$ and $P_4$ are defined in (\ref{lowcorre}).
We carry out a  linear level analysis with $b_1$ only,
and find a real zero with the smallest absolute value at
\begin{eqnarray}
g_{c} \simeq -0.0982 \,,
\end{eqnarray}
and the corresponding exponent is
$\gamma \simeq 2.40$, which is curiously close to the exact value.
These values have $17.9$\% error for $g_c$ and $3.99$\% error for
$\gamma$.\\

In summary, in the lower order calculation,
we have qualitatively good results.
After including higher order perturbative corrections 
and applying Schwinger-Dyson equations, we have achived
some quantitative improvement of the approximated values of the fixed point.
The nontrivial saddle point of $\alpha$ for the effective action
$V[\phi,\alpha]$ is not relevant as far as we have
used lower order results, but it might become important if we include
higher order corrections, since it is capable of reproducing
much nicer numerical results.


\section{The Yang-Mills Matrix Model}
\label{sec:yang-mills-matrix}

In this section, we study the matrix RG of the YM matrix model with
mass terms (\ref{YMmm}).
For convenience of later calculation, 
we rescale the matrices as
 $m\hat{A}^2\rightarrow \hat{A}^2$ and $M\hat{B}^2\rightarrow
 \hat{B}^2$.
The matrix model action is now
\begin{eqnarray}
\frac{1}{2}N \tr \left(\hat{A}^2+\hat{B}^2-\lambda[\hat{A},\hat{B}]^2\right)+
  \frac{1}{2}N^2 \log (mM) , \qquad \lambda \equiv\frac{g}{mM} \,,
\end{eqnarray}
and the matrices $\hat{A}$ and $\hat{B}$ are decomposed as
\begin{eqnarray*}
&&\hat{A}=\left(\begin{array}{cc}
A &a \\
a^{\dagger} & \alpha 
 \end{array}\right), \qquad 
\hat{B}=\left(\begin{array}{cc}
B & b \\
b^{\dagger} & \beta 
\end{array}\right) \,.
\end{eqnarray*}
We integrate out $a$, $a^{\dagger}$, $b$, $b^{\dagger}$, $\alpha$, and
$\beta$ 
to obtain an effective action for $(N-1) \times (N-1)$ matrices.
Under this decomposition, the matrix model action is 
\begin{eqnarray}
NS[\hat{A},\hat{B}]= NS[A,B]+\frac{1}{2}N^2\log(mM)
+\frac{1}{2}N 
\tr\left(\alpha^2+\beta^2\right)
+N v^{\dagger}\Upsilon v
+N\lambda V_4, 
\end{eqnarray}
\begin{align}
v=& (a,b), \qquad 
V_4= |a|^2|b|^2+(a^{\dagger}\cdot b)(b^{\dagger}\cdot a)-
(a^{\dagger}\cdot b)(a^{\dagger}\cdot b)-(b^{\dagger}\cdot
a)(b^{\dagger}\cdot a)  \,,
\nn\\
\Upsilon=&
\left(\begin{array}{cc}
{\bf 1} & 0 \\ 0 & {\bf 1}
\end{array}\right)
+\lambda 
\left(\begin{array}{cc}
B^2+2\beta B+\beta^2{\bf 1}&AB-2BA+\beta A+\alpha B-\alpha\beta {\bf 1} \\
BA-2AB+\alpha B+\beta A-\alpha\beta{\bf 1}&A^2+2\alpha A+\alpha^2{\bf 1}
\end{array}\right) \,.
\label{yyvertices}
\end{align}
The propagators are
\begin{eqnarray}
\langle v_A\;v^{\dagger}_B\rangle_0 =\frac{1}{N}(\Upsilon^{-1})_{AB}, 
\quad \langle \alpha \alpha\rangle_0 =\langle \beta \beta\rangle_0=\frac{1}{N},
\end{eqnarray} 
where  $v_A=a$ and $v_B=b$.

\subsection{Lower order calculations}
First, we carry out the coarse-graining by perturbation theory
up to the second order in $\lambda$. 
In perturbative expansion, $\alpha$ and $\beta$ give sub-leading
effects in the large-$N$ limit, and we set $\alpha=\beta=0$ and do not
include their contributions.
The perturbative calculation gives
\begin{align}
NS[\hat{A},\hat{B}]
 \rightarrow &\;
\frac{1}{2}N \tr \left(A^2+B^2-\lambda[A,B]^2 \right)
+\frac{1}{2}N^2 \log(mM)
+N P_{00}(\lambda) \nonumber \\
&
+P_{20}(\lambda)\tr A^2+P_{02}(\lambda)\tr B^2  
+\lambda P_{22}^{-}(\lambda) \tr [A,B]^2
\nn\\&
+\lambda P_{40}(\lambda) \tr A^4+\lambda P_{04}(\lambda)\tr B^4
+\lambda P_{22}^{+}(\lambda)\tr \{A,B\}^2,
\end{align}
where
\begin{eqnarray}
&&P_{00}(\lambda)=P_{20}(\lambda)=P_{02}(\lambda)=\lambda-\lambda^2, \nonumber \\
&&P_{40}(\lambda)=P_{04}(\lambda)=-\frac{\lambda}{2}, \quad
P_{22}^{-}(\lambda)=\frac{9}{4}\lambda, \quad P_{22}^{+}(\lambda)=-\frac{\lambda}{4}.
\label{2ndPyy}
\end{eqnarray}
It should be noted that we only keep the leading order contributions
in the large-$N$ limit.
To this order, the terms that are absent in the original action,
$\tr A^4$, $\tr B^4$ and $\tr \{A,B\}^2=\tr(AB+BA)^2$,  are induced. 
For the time being, we discard these terms and derive
the beta function of $\lambda$.
Later, we use Schwinger-Dyson equations to eliminate some of them.
The method is parallel to the case of the $\phi^4$ model in the previous
section.
The rescaling of matrices are chosen as
\begin{eqnarray}
\frac{N}{2}\left(1+\frac{2 P_{20}}{N}\right)\tr A^2\rightarrow
\frac{N-1}{2}\tr A^2, \qquad 
\frac{N}{2}\left(1+\frac{2 P_{02}}{N}\right)\tr B^2\rightarrow
\frac{N-1}{2}\tr B^2 .
\end{eqnarray}
We obtain the beta function
\begin{align}
\beta(\lambda)=& -\lambda \big( 1+2P_{20}+2P_{02}+2P_{22}^{-} \big)
\nn\\=&
-\lambda\left(1+\frac{17}{2}\lambda-4\lambda^2\right) \,,
\label{lambdabeta}
\end{align}
and a fixed point and the corresponding critical exponent
\begin{eqnarray}
\lambda_c \sim -0.112,  \qquad
\gamma \sim 1.90 \,.
\end{eqnarray} 
This result implies that the YM-type matrix model (\ref{YMmm}) would
develop
 a non-analytic behavior (\ref{Fsing}) around a certain negative
 $\lambda$ and that
it would be similar to the one from the $\phi^4$ matrix model.

\subsection{Higher order corrections}

We have considered the matrix RG by use of naive perturbation theory
to $\mathcal{O}(\lambda^2)$.
 In this subsection, we consider to include higher order effects
to improve our previous result. 
As in the case of the $\phi^4$ model, we will explore
exact propagators, Schwinger-Dyson equations, and
the effects of nontrivial saddle points of $\alpha$ and $\beta$,
in order.

\paragraph{Exact propagators:}
Let $\frac{1}{N}C_{AB}$ be the exact propagator $\vev{v_{A}
  v^{\dagger}_B}$.
They satisfy the following recursive condition,
\begin{eqnarray}
\frac{C_{AB}}{N}=\frac{\Upsilon^{-1}_{AB}}{N}+\frac{\Upsilon^{-1}_{AC}}{N}
D^{1PI}_{CD}\frac{C_{DB}}{N} .
\label{yypsd}
\end{eqnarray}
These equations iteratively reproduce the perturbative expansions of the propagators. 
In (\ref{yypsd}), $\frac{1}{N}\Upsilon^{-1}_{AB}$ are tree level
propagators and $D^{1PI}_{AB}$ are contributions from 1PI graphs
that are calculated by the exact propagators themselves with the interaction vertex $V_4$. 
In the large-$N$ limit, the leading part of $D^{1PI}_{AB}$ are simply
\begin{align}
D^{1PI}_{aa}=&
-\lambda \tr C_{bb} \cdot \mathbf{1} , \qquad
D^{1PI}_{ab} = -\lambda \left( \tr C_{ba} -2  \tr C_{ab}\right) \mathbf{1}
\,,
\nn\\
D^{1PI}_{bb}=& -\lambda \tr C_{aa} \cdot \mathbf{1} , \qquad
D^{1PI}_{ba} = -\lambda \left( \tr C_{ab} -2 \tr C_{ba}\right)\mathbf{1}  .
\label{eq:10}
\end{align}
From \eqref{eq:10} and (\ref{yypsd}), we can determine the exact
propagators.
In general, it is difficult to solve the  matrix equations (\ref{yypsd}) exactly. 
In the case of $A=B=0$ and $\beta=0$ (or $\alpha=0$), 
we can find a solution, as before.
The details are presented in Appendix \ref{sec:an-exact-solution}
and the solutions are, by setting $C_{AB}=c_{AB}{\bf 1}$,
\begin{eqnarray}
&&c_{aa}=\frac{1+\lambda\alpha^2}{2\lambda}\left(
 -1+\sqrt{1+\frac{4\lambda}{1+\lambda\alpha^2}} \right), \quad  
c_{bb}=\frac{1}{2\lambda}
\left(-1+\sqrt{1+\frac{4\lambda}{1+\lambda\alpha^2}}\right), \nonumber \\
&&c_{ab}=c_{ba}=0.
\label{yyexp}
\end{eqnarray}
These propagators are again partially-resummed propagators
that
include all corrections from the vertex $V_4$ as well as $v$--$\alpha$
four point interactions, but none of corrections involving $A$, $B$
and $\beta$.

Now we consider a simple case with $\alpha=0$ and define $c=c_{aa}=c_{bb}=
\frac{1}{2\lambda}(-1+\sqrt{1+4\lambda})$.
With these propagators,
the corrections to the second order perturbation (\ref{2ndPyy}) can be 
obtained by replacing the coefficients with
\begin{eqnarray}
&&
P_{20}=P_{02}=\lambda c,
\quad
P_{40}=P_{04}=-\frac{\lambda}{2} c^2, \quad 
P_{22}^{-}=\frac{9}{4}\lambda c^2, \quad P_{22}^{+}=-\frac{\lambda}{4}c^2.
\label{exPyy}
\end{eqnarray}
In this case, the fixed point of the beta function and the corresponding
exponent are calculated to be
\begin{eqnarray}
\lambda_c\simeq -0.0982, \qquad \gamma\simeq 1.64.
\end{eqnarray}

\paragraph{Schwinger-Dyson equations:} 

We have seen that the perturbative calculation to the second order
induces $\tr A^4$, $\tr B^4$ and $\tr\{A, B\}^2$ terms that do not
exist in the original action \eqref{YMmm}.
As in the case of the $\phi^4$ model,
 we use the Schwinger-Dyson equations to replace some of the 
induced operators with other operators and investigate the change of
the couplings.

The set of the relations we need are worked out in Appendix
\ref{sec:schw-dyson-equat}, and the result is
\begin{eqnarray}
\vev{ \tr A^4}= \vev{ \tr B^4 }
= -\frac{1}{4} \vev{\tr [A,B]^2 }
+\frac{3}{4}\vev{\tr \{ A,B\}^2} \,.
\end{eqnarray}
With $\tr A^4$ and $\tr B^4$ terms replaced by use of these relations,
(\ref{2ndPyy}) is modified as
\begin{eqnarray}
&&P_{00}=P_{20}=P_{02}=\lambda-\lambda^2,
\quad
P_{40}=P_{04}=0, \quad
P_{22}^{-}=\frac{5}{2}\lambda, \quad P_{22}^{+}=-\lambda.
\label{2ndPyy+sd}
\end{eqnarray} 
Using  them,  we get
\begin{eqnarray}
\lambda_c
\simeq -0.106 \,,
 \qquad \gamma
\simeq 1.91 \,.
\end{eqnarray}

\paragraph{Effects of nontrivial saddle points of $\alpha$ and $\beta$:}

Here, we discuss whether a non-trivial saddle point of $\alpha$ and $\beta$ integrals changes the lower order results.

Just as in the case of the $\phi^4$ matrix model,
the perturbative calculation to $\mathcal{O}(\lambda^2)$ 
gives rise to the following effective action including $\alpha$ and $\beta$,
\begin{align}
V[A,B;\alpha,\beta]= &
V[A,B;0,0]+
\frac{1}{2}\left(1+2(\lambda-\lambda^2)\right)(\alpha^2+\beta^2)
\nonumber \\ &
-\frac{\lambda^2}{2}\left((\alpha^2+\beta^2)^2+2\frac{\tr (A^2+B^2)}{N}(\alpha^2+\beta^2)
+4\frac{\tr A^2}{N}\alpha^2+4 \frac{\tr B^2}{N}\beta^2
\right)
\,,
\end{align}
where we have dropped the terms
proportional to $\tr A,\tr B, \tr AB, \tr A^3, \tr B^3, \tr A^2B$, and $\tr AB^2$ which
will vanish after the averaging with respect to $S[A,B]$.
We adopt a parametrization
\begin{eqnarray}
\alpha=u\cos\theta, \quad \beta=u\sin\theta,
\end{eqnarray}
and first consider the saddle point with respect to $\theta$,
$\frac{\partial V}{\partial \theta}=0$.
Since all the solutions, $\theta=0,\frac{\pi}{2}, \pi$, and
$\frac{3\pi}{2}$,
serve essentially an equivalent result,
we take $\theta=0$ for further study. 
Now the effective potential becomes
\begin{align}
V[A,B;u , \theta=0]=
V[A,B;0,0]+
\frac{1}{2}\left(1+2(\lambda-\lambda^2)-2\lambda^2\frac{3 \tr A^2+ \tr
    B^2}{N}\right)u^2-\frac{\lambda^2u^4}{2}.
\label{upoten}
\end{align}
There is no non-trivial saddle point which gives an energy
lower than $V[A,B; 0,0]$.
Here we provide a suggestion to include some of higher-order corrections and observe if there
appear nontrivial saddle points.

In (\ref{yyvertices}), the Gaussian integration of $v,v^{\dagger}$ can
be done if $A=B=0$ and $V_4$ interaction is neglected.
This gives rise to an effective potential
\begin{eqnarray}
V[A,B; u,\theta] = V[A,B;0,0]+\frac{u^2}{2}+\log(1+\lambda u^2).
\end{eqnarray}      
Next, we introduce first order contributions of $A^2$ and $B^2$.
We guess that, for $\theta=0$ case,
\begin{eqnarray}
V[A,B;0,0]+\frac{1}{2}\left(1-2\lambda^2-2\lambda^2\frac{\tr (3A^2+B^2)}{N}\right)u^2+\log(1+\lambda u^2).
\label{upoten2}
\end{eqnarray} 
The potential (\ref{upoten2}) agrees with the original potential
(\ref{upoten})
to $O(\lambda^2)$.
Using  the potential (\ref{upoten2}), we consider  the saddle point equation
 \begin{eqnarray}
0=u\left[ \left(1-2\lambda^2-2\lambda^2\frac{\tr (3A^2+B^2)}{N}\right)+\frac{2\lambda }{1+\lambda u^2}\right].
\end{eqnarray} 
A non-trivial solution is
\begin{eqnarray}
u^2_c=\frac{-1}{\lambda}-\frac{2}{1-2\lambda^2-2\lambda^2\frac{\tr (3A^2+B^2)}{N}}.
\end{eqnarray}
The reality condition for $u_c$, $u_c^2\ge 0$, is evaluated as
$\lambda \le -0.316$,  $-0.232\le \lambda\le 0$ or $0.316 \leq \lambda \leq 0.432$, using
$\frac{1}{N}\langle\mbox{tr}A^2\rangle=1+O(\lambda)$ and
$\frac{1}{N}\langle \mbox{tr}B^2\rangle=1+O(\lambda)$.
The condition for the effective potential evaluated by this saddle
point to be real leads to further conditions on $\lambda$,
and the allowed region is evaluated as
$-0.232 \leq \lambda \leq  0$ or $0.316 \leq \lambda \leq 0.432$.
The difference of the potential values at the saddle point
$V[u_c]-V[0]$ turns out to be
positive semi-definite in this allowed region,
 and the saddle points do not contribute like
the case of the $\phi^4$ model.
Despite of this observation, we again try to evaluate the
values of the fixed point and the corresponding exponent,
associated with this non-trivial saddle point.
We expand the saddle point action $V[A,B;u_c,0]$
by $\frac{1}{N}\mbox{tr}(3A^2+B^2)$ term\footnote{It imposes an
  additional reality condition, $\frac{2\lambda}{2\lambda^2-1}>0$.}
 to the first order.
This produces a new correction term
\begin{eqnarray}
\kappa(\lambda) \frac{\tr (3A^2+B^2)}{N},
\qquad
\kappa(\lambda) = \frac{\lambda(1+2\lambda-2\lambda^2)}{1-2\lambda^2} \,,
\end{eqnarray} 
which modifies the beta function as
\begin{eqnarray}
\beta(\lambda)= -\lambda \left(
1+ 2\left(P_{20}+3\kappa(\lambda) \right)
+2\left( P_{02} + \kappa(\lambda) \right) 
+2P^{-}_{22}  \right).
\end{eqnarray}
Plugging (\ref{2ndPyy+sd}) into this, a fixed point and the
corresponding
critical exponent are evaluated as
\begin{eqnarray}
\lambda_c\simeq -0.0615, \quad \gamma\simeq 2.10.
\end{eqnarray}
We have got a suggestive result from the above analysis,
like the case of the $\phi^4$ model.

The saddle point here turns out to not contribute to the matrix RG
in the large-$N$ limit,
but neglecting $V_4$ interaction is not a good approximation,
as this self interactions are comparable to the $v$--$\alpha$,
$v$--$\beta$ interactions.
In order to take account of the effect of the self-interaction,
we next consider to use the exact propagators (\ref{yyexp}) which are valid when
$A=B=\beta=0$.
We employ them to integrate over $v$ and $v^{\dagger}$, 
and take the lowest order corrections from $A$ and $B$.
Then, we find an effective potential
\begin{align}
  \label{eq:13}
  V[A,B;\alpha,0]=&
\tilde{V}[A,B;0,0]
+\frac{\alpha^2}{2} 
\nn\\&
+ \lambda \left(
\frac{c_{bb}}{N} \tr A^2 + \frac{c_{aa}}{N} \tr B^2 \right)
-\log c_{aa} - \log c_{bb} - \lambda c_{aa} c_{bb} \,,
\end{align}
where in the second line, $\alpha=0$ part contains
$\alpha$-independent corrections, and $\tilde{V}[A,B;0,0]$ means that
these corrections are subtracted from $V[A,B;0,0]$.
This potential is still too complicated to extract a tractable saddle point.
We may try to find a nonperturbative saddle point of $\alpha$ in the
weak coupling regime, $\lambda \rightarrow 0$ with $\lambda \alpha^2$
fixed.
In this case, $c_{bb} \rightarrow (1+\lambda\alpha^2)^{-1}$
and $c_{aa} \rightarrow 1$, and
\begin{align}
  \label{eq:14}
  V[A,B;\alpha,0] \rightarrow \tilde{V}[A,B;0,0]
+\frac{\lambda}{N} \tr B^2 + \frac{\alpha^2}{2}
-\frac{\lambda}{1+\lambda \alpha^2} \left(1- \frac{\tr A^2}{N} 
\right)
+ \log (1+\lambda \alpha^2) \,. 
\end{align}
Nontrivial saddle points appear at
\begin{align}
  \label{eq:15}
  X_\pm \equiv \frac{2 \lambda}{1+\lambda \alpha^2_\pm} = 
Y^{-1} (-1 \pm \sqrt{1-2Y}) \,,
\qquad
Y=1- \frac{\tr A^2}{N} \,.
\end{align}
In the weak coupling regime, we may evaluate $Y$ by use of
perturbation theory, $\frac{1}{N} \vev{\tr A^2}=1-2\lambda
+12\lambda^2 + \mathcal{O}(\lambda^3)$,
as $Y=2\lambda -12\lambda^2 + \mathcal{O}(\lambda^3)$.
Thus, $X_\pm$ will be $X_+=-1-\lambda +4\lambda^2 + \mathcal{O}(\lambda^3)$ for the
positive sign solution and $X_-=-\lambda^{-1}-5+\mathcal{O}(\lambda)$
for the negative one.
In terms of $\alpha$, the saddle points are
\begin{align}
  \label{eq:16}
  \alpha^2_\pm =& -\frac{1}{\lambda} + \frac{2}{X_\pm} \,.
\end{align}
This is self-consistent as $\alpha^2_\pm \lambda =-1 + \mathcal{O}(\lambda)$.
The difference of the potential is
\begin{align}
  \label{eq:17}
  V[A,B;\alpha_\pm,0] - V[A,B;0,0]=&
\frac{\alpha_\pm^2}{2} +
\frac{\lambda^2 \alpha_\pm^2}{1+\lambda\alpha_\pm^2} Y + \log
(1+\lambda \alpha_\pm^2) \,.
\end{align}
In order for the potential to remain real, we need to take the $X_+$
solution and also keep $\lambda$
to be a small negative value.
This also retains $\alpha^2_+$ to be positive.
If we keep up to $\mathcal{O}(\lambda^2)$ terms in the small $\lambda$ expansion as
well as the logarithmic term, by use of the above perturbative estimation,
we find that the difference is always positive.
In this second analysis,  the nontrivial fixed points would not contribute to
the effective potential at least within the region
the analysis is valid.\\

In the previous and this subsections, we have analyzed the matrix RG equation
for the Yang-Mills type matrix model with the mass terms.
We derive the beta function in various ways
and investigate the fixed points and the corresponding
exponents.
The result suggests that the model would exhibit a similar critical behavior
to that of the $\phi^4$ model for a small negative value of $\lambda=g/Mm$.
In parallel with the case of the $\phi^4$ model, 
we have examined the effects from non-trivial
saddle points of an effective potential of $\alpha$ and $\beta$.
After we investigate that
perturbative calculations do not lead to a non-trivial saddle point, 
we try to consider a partially resummed potential,
based on a Gaussian integration as well as
using the exact propagators.
These two arguments provide different observations,
and
we have not found a conclusive evidence that
a non-trivial saddle point of the effective
$\alpha$ and $\beta$ potential has
a significant effect to evaluate the exponent
$\gamma$.
It might be the case if we could include higher order effects
in a more systematic way,
but a more detailed investigation is left
as a future work. 
As for the analysis based on the Schwinger-Dyson equations,
on contrary to the $\phi^4$ model, we cannot eliminate all the
unwanted terms, such as $\tr \{ A,B\}^2$.
Obviously, the action \eqref{YMmm} is not the most
general one respecting all the symmetries it possesses,
and it may not be surprising that we cannot control the higher order
correction by only using the Schwinger-Dyson equations.
On the other hand, in \eqref{Ymmm2}, the SD equation analysis might give a closed
form, but we do not carry out this study in the present paper.

\subsection{A critical behavior of the Yang-Mills matrix model with the
  mass terms}

In the study of the YM two matrix model with the mass terms,
we can diagonalize one of
the matrices to derive a saddle-point equation for the eigenvalue distribution
of the diagonalized matrix.
An analytic solution has been discussed in the literature, and we discuss
its implication to our result.
Another qualitative discussion about
the critical behavior for of the YM matrix model with the mass terms
is presented in
Appendix \ref{sec:matrix-analogue-mean}, based on a matrix analogue of
the mean field approximation.

\paragraph{An implication from exact results:}

We can study the YM two-matrix models by diagonalizing one of the two
matrices.
We  choose to diagonalize $\hat{A}$ in (\ref{YMmm}). 
The action becomes 
\begin{eqnarray}
\hspace{-5mm}
S=\frac{1}{2} \left( \sum_{i=1}^{N}(m\lambda_i^2+M\hat{B}_{ii}^2)
+\sum_{i\neq j}^{N}\{M+g(\lambda_i-\lambda_j)^2\}\hat{B}_{ij}\hat{B}_{ji} \right)
-\frac{1}{N}\sum_{i\neq j}\log|\lambda_i-\lambda_j|,
\end{eqnarray} 
where $\lambda_i$ are the eigenvalues  of the matrix $\hat{A}$,
and the last log term comes from the Vandermonde determinant.
The Gaussian integration with respect to
$\hat{B}$ leads to
\begin{align}
Z \propto &\prod_{i=1}^{N}\int d\lambda_i\;e^{-NV},
\nn\\ &
V=
\sum_{i=0}^{N}\frac{m}{2}\lambda_i^2
+\frac{1}{N}\sum_{i < j}\Big(\log \{M+g(\lambda_i-\lambda_j)^2\}-\log(\lambda_i-\lambda_j)^2 \Big).
\end{align}
In the large-$N$ limit,
the solution of the eigenvalue distribution may be obtained by solving
the following saddle point equations,
\begin{eqnarray}
m\lambda_i+\frac{1}{2 N}\sum_{j(\neq i)}\left\{
\left(\frac{1}{\lambda_i-\lambda_j-i\sqrt{\frac{M}{g}}}-\frac{1}{\lambda_i-\lambda_j}\right)
+\left(\frac{1}{\lambda_i-\lambda_j+i\sqrt{\frac{M}{g}} }-\frac{1}{\lambda_i-\lambda_j}\right)
\right\}=0. 
\nonumber \\
\label{yylNsadl}
\end{eqnarray} 
Looking at the  matrix integral after $\hat{B}$ is integrated out,
one finds that this is similar to the Gaussian matrix model when $g$
is small.
Therefore, we expect that there is a large-$N$ solution with a single brunch
cut.
If $g$ is negative, the log potential becomes unbounded from below
and a critical behavior is expected to come in.

The saddle point equation \eqref{yylNsadl} is solved in
\cite{Kazakov:1998ji}
as implicit functions of the coupling constant and the free energy
in the large-$N$ limit.
 The solution is written in terms of the standard elliptic functions,
and it is not straightforward to extract critical behaviors from the
solution.
\cite{Kazakov:1998ji} argued that that the singular behavior of the
planar part of the free energy should be that of pure two dimensional
gravity.
The sketch of the argument is in the following.
In the commutator interaction $\tr [\hat{A},\hat{B}]^2 =
2\tr (\hat{A}^2 \hat{B}^2 - \hat{A}\hat{B}\hat{A}\hat{B})$,
the latter term serves a nonplanar contraction of the matrices,
and then the planar free energy must contain an even number of this
vertex.
Therefore, in the planar limit, one can flip the sign of the latter
interaction,
and it becomes a matrix model studied in
\cite{Eynard:1998}, a three
matrix model for a three-color problem on a random surface.
It is argued that if the critical behavior of this model is realized
as a scaling behavior of the eigenvalue distribution in the vicinity
of the endpoints of its support, it is equivalent to that of
the $O(n)$ loop gas model on a random lattice with $n=1$.
The $O(1)$ loop gas model is known to belong to
the universality class of the pure two dimensional quantum
gravity \cite{Gaudin:1989vx};
namely, that with the critical exponent $\gamma=5/2$.
This argument suggests that
the critical exponent of the YM matrix model with the mass terms,
in the planar limit, is also $\gamma=5/2$.

We may also refer to a discussion by use of the planar analytic
solutions given in \cite{Kazakov:1998ji}.
They are given in a complicated form using the elliptic functions and are not easy to
read off the critical behavior.
However,  the solutions can be simplified by the strong coupling
expansion, which is far beyond the convergence radius of $g$-series.
Let $\nu \equiv \langle \frac{tr\hat{A}^2}{2N}\rangle$
(the factor $2$ comes from the normalization of the matrices).
The solutions become\footnote{Our
  $g$ corresponds to $\frac{g^2}{2}$ of \cite{Kazakov:1998ji}.}
\begin{eqnarray}
g&=& \frac{1}{24\pi^4L^3}(1-3L),
\label{kp01}\\
\nu&=&\frac{1}{20\pi^2L^2}\left(\frac{1-10L+20L^2}{1-3L}\right)+\frac{1}{12} \,.
\label{kp02}
\end{eqnarray}
$L$ is an implicit parameter, and $g \rightarrow \infty$ corresponds to $L \rightarrow 0$.
In these expression, higher order terms in $L$ are dropped ($L$ is
written as the logarithm of another parameter and this is called the
leading logarithmic approximation in \cite{Kazakov:1998ji}).
The authors of \cite{Kazakov:1998ji} also make an interesting remark
that the above equations are equivalent to the genus-zero part of a KP
equation and the grand partition function of the YM matrix model is
also a specific tau-function of the KP hierarchy. 
Now let us solve the first equation in (\ref{kp01}) with respect to
$L$.
We find one real root and two complex roots, and the real one is 
\begin{eqnarray}
L=\frac{
6^{\frac{1}{3}}\left(
g^{\frac{4}{3}}\left(6\pi^2+\sqrt{6}\sqrt{\frac{1+6\pi^4g}{g}}\right)^{\frac{2}{3}}
-6^{\frac{1}{3}}g
\right)}{12\pi^2g^{\frac{5}{3}}
\left(6\pi^2+\sqrt{6}\sqrt{\frac{1+6\pi^4g}{g} }
\right)^{\frac{1}{3}}}.
\end{eqnarray}
We find a branch point at $g_c=-\frac{1}{6\pi^4}$.
Plugging this expression into  (\ref{kp02}) and expand it around
$\sqrt{g}=\sqrt{g_c}$ (our coupling constant corresponds to $-g^2$ in
\cite{Eynard:1998}), one can check the singular behavior of $\nu$ as
$\nu\sim (\sqrt{g}-\sqrt{g}_c)^{\frac{3}{2}}$, which agrees with that
of 2D pure quantum gravity.
However, $\nu$ is now imaginary.
If we want to avoid an imaginary $\nu$, one may choose a suitable
branch  $(-1)^{1/3}=-1$
and expand $\nu$ with respect to $g-g_c$. 
In this case, we find no singular behavior in $L$ and $\nu$ around $g\sim g_c$.
It should be noted that, as already pointed out in
\cite{Kazakov:1998ji},
$g_c=-\frac{1}{6\pi^4}$ is the point where the leading logarithmic
approximation is not really valid, and also this analysis is based on
the large-$g$ expansion, while our matrix RG analysis is trustable when
$g$ is sufficiently small.
\\

We have examined 
the arguments that
 the YM matrix model with the mass terms will develop
a critical behavior for a negative value of $g$, referring to an
analytic work \cite{Kazakov:1998ji}.
Together with the discussion of the mean field approximation in
Appendix \ref{sec:matrix-analogue-mean}, they provide supports that
the matrix RG analysis for the YM matrix model captures this critical
behavior, at least, qualitatively.

\section{The Massless Limit}
\label{sec:massless-limit}

In this section, we consider the massless limit of the matrix RG
analysis.
Under the massless limit $m,M \rightarrow 0$, the YM matrix models we have considered
\eqref{YMmm} becomes the usual large-$N$ reduced model of
pure Yang-Mills theory in two dimensions.
This model is a two dimensional counterpart of the bosonic IKKT model,
and the understanding of its nonperturbative dynamics is of particular
interest
and is expected to be a touchstone toward the understanding of the
supersymmetric model and superstring theory.

As well known, the two dimensional bosonic IKKT matrix model is ill-defined,
as its partition function is divergent for arbitrary $N$.
It has been shown that when the dimension is larger than two, namely
there are three or more matrix fields, the partition function is
convergent
for sufficiently large and fixed $N$ \cite{Krauth:1999qw, Austing:2001bd}.
In this paper, we take our two dimensional model as a toy model for
these more well defined ones, and simply neglect the divergence due to
the massless limit.
In short, we start with the massive model and consider the massless
limit of the beta functions afterwards.
We expect that this prescription still captures an essential behavior
of general YM matrix model under the massless limit.

Another point to be noted is that after dropping the quadratic terms
in \eqref{YMmm}, the only remaining parameter $g$ can be scaled out
and is understood as a scale parameter.
Namely, the theory is parameterless, which is also true for its
supersymmetric extensions.
Therefore, there is no ``critical point'' with respect to the coupling
constant
in this model.
We, instead, look at the generation of the quadratic term in the
massless model.
Under the massless limit, the model enjoys a shift symmetry,
$\hat{A} \rightarrow \hat{A} + c \mathbf{1}$ and
$\hat{B} \rightarrow \hat{B} + c' \mathbf{1}$, where
$\mathbf{1}$ is a unit matrix and $c,c'$ are arbitrary constants.
Therefore, one may think that the expectation value
of an operator which does not respect this symmetry, such as
$\vev{\tr \hat{A}^2}$, is going to vanish.
On the other hand, it is also known that the effective potential for
the diagonal elements of the matrices, which is generated by
integrating out the off-diagonal elements, exhibits
repulsive behavior when these two elements are close to each other\footnote{%
In the YM two matrix model in question, this analysis should be done with
small mass parameters, and the behavior of the eigenvalues are examined under
the massless limit.
The long-range, compared to the regulator scale,
attractive behavior is gone for the two dimensional case
and the theory suffers an IR divergence.
The short distance behavior may be valid under a tuned massless
limit discussed here.}\cite{Hotta:1998en}.
 Therefore the diagonal elements tend to spread out and to generate
nonvanishing values of, for example, the quadratic operator above.
We will discuss that the matrix RG analysis captures this feature in a massless limit.

Through the procedure of the matrix RG, the change of the action for \eqref{YMmm}
is (now the mass parameters are restored)
\begin{eqnarray}
S
\rightarrow \frac{1}{2}
\tr\left( m\left( 1+\frac{2P_{20}}{m N} \right) A^2+M \left( 1+\frac{2P_{02}}{M N}\right) B^2
-g\left( 1-\frac{2P_{22}^{-}}{N} \right) [A,B]^2+ \cdots  \right).
\label{eq:20}
\end{eqnarray} 
We retain the interaction term, instead of the quadratic terms in the previous
considerations, to be canonical by the rescaling,
$-\frac{gN}{2} \left(1-\frac{2P_{22}^-}{N}\right) \tr[A,B]^2
\rightarrow -\frac{g(N-1)}{2} \tr[A,B]^2$.
Since this term involves both $A$ and $B$, there is a choice of freedom
to compel which of them to be rescaled.
We take a symmetric choice here, and, as we will see, the essential result
does not depend on this choice.
Under this rescaling, the mass terms become
\begin{eqnarray}
\frac{m(N-1)}{2}  \left(1+\frac{1/2+2P_{20}/m+P_{22}^{-}}{N}\right) \tr A^2, 
\frac{M(N-1)}{2} \left(1+\frac{1/2+2P_{02}/M+P_{22}^{-}}{N}\right)\tr B^2. 
\nonumber \\
\end{eqnarray} 
The changes of the mass terms are described by the following beta functions,
\begin{eqnarray}
\beta_m= m\left(\frac{1}{2}+2\frac{P_{20}}{m}+P_{22}^{-}\right), \quad 
\beta_M= M\left(\frac{1}{2}+2\frac{P_{02}}{M}+P_{22}^{-}\right).
\end{eqnarray}
After the massless limit, $m,M\rightarrow 0$, we obtain the beta functions of
the YM model without the mass terms.
Since perturbation theory breaks down under the massless limit, 
one cannot naively take the massless limit in these expressions.
In order to circumvent this difficulty,
we may use the exact propagators $c_{aa}$ and $c_{bb}$ to evaluate these beta functions.
After restoring the mass parameters, the exact propagators are
given by $c_{aa} = c/m$ and $c_{bb}=c/M$ where
 $c=\frac{1}{2\lambda}(-1+\sqrt{1+4\lambda})$ is the one given
in the previous section.
The values of \eqref{exPyy} are replaced as
\begin{align}
  \label{eq:11}
  P_{20}=& g c_{bb} \,,
\quad
P_{02}= g c_{aa} \,,
\quad
P_{22}^-= \frac{9}{4} g c_{aa} c_{bb} \,,
\end{align}
and then
\begin{eqnarray}
\beta_m=m\left(\frac{1}{2}+2\lambda c+\frac{9}{4}\lambda c^2\right), \quad
\beta_M=M\left(\frac{1}{2}+2\lambda c+\frac{9}{4}\lambda c^2\right).
\label{eq:19}
\end{eqnarray} 
We consider a massless limit as $m,M\rightarrow 0$ with the ratio $m/M$ fixed,
which leads to finite results,
\begin{eqnarray}
\beta_m= 2\sqrt{g\frac{m}{M}}, \qquad
\beta_M=2\sqrt{g\frac{M}{m}}.
\label{IKKTbetas}
\end{eqnarray} 
In this limit, one can see that only the second terms of each expressions
in \eqref{eq:19} survives.
The second terms come from the corrections in \eqref{eq:20}, while
the first and the third terms are due to the rescaling.
Therefore in this limit, the choice of the freedom of the rescaling is not relevant
for the result.
These beta functions suggest that once the quadratic terms are induced in a certain way,
 they are prone to increase.
This can be understood as a manifestation of the generation of the nonvanishing
expectation value of a quadratic operator $\vev{\tr A^2}$ discussed above.

\paragraph{An improved perturbation method:} 

We move on to considering another approach to deal with the massless YM matrix model,
an improved perturbation method (IPM).
We start with a brief introduction of the IPM 
(also known as a Gaussian approximation or an improved mean field
approximation \cite{Kabat:1999hp,Oda:2000im}).
Let us write the action of our (massive) model as $S[m,M,g]$ to show the dependence
on the parameters explicitly. 
We then consider a deformed action $S[m_0 +x(m-m_0), M_0 +x (M-M_0), gx]$,
where
$m_0$ and $M_0$ are ``mean fields'' to be tuned appropriately as explained soon\footnote{
These ``mean fields'' should be distinguished from the $\sigma$s
discussed in Appendix \ref{sec:matrix-analogue-mean}. They are
different objects. },
$x$ is a parameter to be set to $1$ after a calculation.
It is immediate to see that if we set $x=1$ in this deformed action, the action comes
back to the original one, and $m_0$ and $M_0$ take no effect.
Now we take $x$ as a formal expansion parameter, with which we carry out a perturbative
expansion to some order.
This perturbation series, after setting $x=1$, would depend on the parameters
$m_0$ and $M_0$ as it is a finite order approximation.
Since the exact answer from the original action does not depend on these parameters,
we need to vary $m_0$ and $M_0$ to seek for the point where the approximated value
is not sensitive to the change of them.
So the criterion for the approximation to work well is characterized by
the appearance of the
``plateau'' of the graph for a physical quantity with respect to
 these mean field parameters \cite{Kawai:2002ub}.
To lower orders, a clear plateau would not be formed, and the approximation 
scheme is to take extrema of a physical quantity with respect to
the mean fields.
This method has been extensively applied to IKKT-type matrix models
 to explore the spontaneous breakdown
of the rotational symmetry \cite{Kawai:2002ub, Nishimura:2001sx,
  Nishimura:2004ts,Aoyama:2006di}.

What we would like to investigate is a massless YM matrix model, $S[0,0,g]$.
In this case, the deformed action reads $S[(1-x)m_0, (1-x)M_0, xg]$.
So, practically, we can start with a massive action $S[m_0, M_0, g]$ and calculate
quantities we want, by perturbation theory to some orders, such as the beta functions
and the planar free energy
\begin{eqnarray}
\beta_m= m_0 \left(\frac{1}{2}+2\frac{P_{20}}{m_0}+P_{22}^{-}\right), \quad 
\beta_M= M_0 \left(\frac{1}{2}+2\frac{P_{02}}{M_0}+P_{22}^{-}\right) \,,
\end{eqnarray}
and
\begin{eqnarray}
F_0 = \frac{1}{2} \log ( M_0m_0) + \lambda_0 - 3 \lambda_0^2 + \mathcal{O}(g^3) \,,
\end{eqnarray}  
where $\lambda_0 = g/m_0M_0$.
We then make a replacement
\begin{align}
  \label{eq:21}
  m_0 \rightarrow (1-x) m_0\,, \quad
M_0 \rightarrow (1-x)M_0 \,, \quad
g \rightarrow xg \,,
\end{align}
and expand the above quantities with respect to $x$ to a certain order and set $x=1$.
Here, we demonstrate the IPM at the first order.
The result \eqref{2ndPyy+sd}, which takes the Schwinger-Dyson equations into account,
now reads
\begin{align}
  \label{eq:23}
  P_{20}= m_0 (\lambda_0-\lambda_0^2) \,,
\quad P_{02}=M_0(\lambda_0-\lambda_0^2) \,,
\quad
 P_{22}^- = \frac{5}{2}\lambda_0 \,.
\end{align}
Using them we find the first order improved quantities that depend on the mean fields,
\begin{gather}
  \label{eq:24}
  \beta_m^{(\text{imp})} = \frac{9}{2}m_0\lambda_0 \,,
\quad
\beta_M^{(\text{imp})} = \frac{9}{2}M_0\lambda_0 \,,
\\
F^{(\text{imp})}_0 = -1+\frac{1}{2}\log(M_0m_0) +\lambda_{0}\,.
\end{gather}

We need to fix $m_0$ and $M_0$ by requiring a physical observable to be on a stationary point
with respect to them.
The expectation value of any gauge invariant operator would be suitable for this purpose,
though its choice may affect the approximated values at lower orders.
We here use the improved free energy to fix our parameters.
It turns out that the stationary point conditions,
\begin{align}
  \label{eq:26}
  \frac{\partial F^{(\text{imp})}_0}{\partial m_0} = 
  \frac{\partial F^{(\text{imp})}_0}{\partial M_0} = 0\,,
\end{align}
do not fix $m_0$ and $M_0$ independently\footnote{%
For $D \geq 3$ pure Yang-Mills reduced models, all the mean fields are chosen to be
an equal value \cite{Nishimura:2001sx}.}
 but fix the product $m_0 M_0$.
To fit the discussion to the previous massless-limit one, we fix the ratio $\omega = M_0/ m_0$ and
solve the stationary condition with respect to $m_0$ as
\begin{align}
  \label{eq:27}
  \lambda_0 = \frac{g}{\omega m_0^2} = \frac{1}{2}  \,.
\end{align}
Recall that $g$ is not a free parameter for the massless theory but
a scale parameter, which we would set to be positive.
It is thus  natural to assume that $m_0$, $M_0$, and $g$ (therefore, $\omega$ as well) 
are all positive and to take the positive branch of the solution for $m_0$.
In that case, the improved beta functions are evaluated as
\begin{align}
  \label{eq:28}
  \beta_m = \frac{9}{4} \sqrt{2g \omega^{-1}} = \frac{9}{2\sqrt{2}} \sqrt{g \frac{m_0}{M_0}} \,,
\qquad
\beta_M= \frac{9}{4} \sqrt{2 g\omega}= \frac{9}{2\sqrt{2}} \sqrt{g\frac{M_0}{m_0}} \,.
\end{align}
Although the parameter ratios $m/M$ and $m_0/M_0$ possess different physical meanings,
these improved beta functions appear in interestingly similar forms to \eqref{IKKTbetas}.
If we carry out IPM to the second order, it turns out that (\ref{eq:26})
has no real solution. However in the third order, (\ref{eq:26}) has at
least one real solution. 
At higher orders, we may also expect a plateau to emerge.

It will be interesting if we can discuss the spontaneous breaking (or non-breaking
for the bosonic models) of the rotational symmetry of IKKT-type matrix models,
by combining the improved perturbation method with
the matrix RG scheme.

\section{Summary and Discussion}
\label{sec:summary-discussion}

In this paper, we study a YM-type two matrix model with  mass
terms by use of the matrix RG approach proposed by Brezin and
Zinn-Justin \cite{Brezin:1992yc}.

In Section \ref{sec:phi4-model}, we first revisit the matrix RG of the
$\phi^4$ matrix model,
which has been discussed in the literature to some extent.
In this model, the analytic solution is available and we can test
how well the RG method works.
A naive application of perturbation theory, as already done in
\cite{Brezin:1992yc}, provides a reasonable but quantitatively
not so satisfactory result.
We employ several methods
to improve the results, which include
an exact propagator, the Schwinger-Dyson equation, and
the search for a non-trivial saddle point of $\phi_{NN}$ potential.
Through them, we have observed various improvement of the approximation.

In Section \ref{sec:yang-mills-matrix}, the YM matrix model with mass
terms is studied.
We use a simple perturbative expansion,
an exact propagator method and the Schwinger-Dyson equations, and also search for a non-trivial saddle point.
We find that the model has somehow similar behavior
to the $\phi^4$ model.
Especially, it indicates that this model exhibits
a critical behavior for a small negative coupling constant.
On the other hand, the large-$N$ exact solution of this model has been
obtained \cite{Kazakov:1998ji}.
One may extract the critical behavior in the planar limit
from this solution, and it 
is discussed to be the same one as the
$\phi^4$ model with the critical exponent $\gamma=5/2$. 
We discuss this would be true under the leading logarithmic
approximation of the analytic result in the large coupling regime \cite{Kazakov:1998ji},
but, there, the free energy turns out to take an imaginary value.

We finally try to discuss the YM model without the mass terms,
by considering the massless limit of the matrix RG equation.
In the massless theory,  there is no critical behavior of the free
energy, and we instead 
examine the beta functions for the quadratic terms in the massless
limit.
This should be related to the dynamical generation of the mass terms 
(the expectation value of the quadratic terms)
of the model, which may be a universal feature of the large-$N$
hermitian matrix models,
and we obtain a consistent answer to these expectations.

The method employed in this paper can be
applied to other YM type matrix models,
either reduced higher-dimensional Yang-Mills theory
or supersymmetric extensions, in a straightforward manner.
Among them, the nonperturbative aspects of IKKT model is of
particular interest, and
we expect that the matrix RG approach sheds more light on
the properties of these matrix models.

Finally, we make a remark on a recent new formulation of the matrix RG by
utilizing the fuzzy sphere structure \cite{Kawamoto:2012ng}.
Such a formulation can be useful to analyze YM-type models without 
quadratic terms.
The YM potential term itself may produce quadratic terms if we consider a
fuzzy sphere background,
and we do not have to introduce the mass terms by hand.

\section*{Acknowledgments}

The work of SK is supported by NSC99--2112--M--029--003--MY3 and NSC101--2811--M--029--001.
The work of DT is supported by  NSC101--2811--M--029--002. 
A part of his work has been done in National Center for Theoretical Sciences (NCTS, Hsinchu).

\appendix

\section{The values of $g_c$ and $\gamma$ from the non-linear RG equation in 
the $\phi^4$ model}
\label{sec:non-linear}

We calculate fixed points of a beta function and the corresponding
critical exponents from the non-linear RG equation (\ref{nlRG}).
We follow the method in \cite{Higuchi:1994rv}.

The singular part of the planar free energy,
which is $N$ independent, can be determined by the RG equation,
\begin{eqnarray}
2F=b_0+b_1\frac{\partial F}{\partial g}+b_2\left(\frac{\partial F}{\partial g}\right)^2.
\end{eqnarray}
We assume the form of a solution near a coupling $g_c$ as
\begin{eqnarray}
F=\sum_{n=0}^{\infty}a_n(g-g_c)^n+\sum_{n=0}^{\infty}d_n(g-g_c)^{\gamma+n},
\label{expF}
\end{eqnarray}
where a non-integral number $\gamma$ describes the 
non-analytic property of $F$ near $g\simeq g_c$,
and $g_c$ should be determined by the RG equation.
We introduce a new variable $a\equiv \frac{\partial F}{\partial g}-a_1$. 
The RG equation becomes
\begin{eqnarray}
2F=\beta_0(g)+\beta_1(g)a+\beta_2(g)a^2, 
\label{nlRG2}
\end{eqnarray}
with
\begin{align}
\beta_{0}(g)=&b_0+a_1b_1+a_1^2b_2, \qquad
\beta_{1}(g)=b_1+2a_1b_2, \qquad
\beta_{2}(g)=b_2.
\end{align}
These $\beta_{n}$ are expanded around $g=g_c$ as
\begin{eqnarray} 
\beta_{n}(g)=\sum_{m=0}^{\infty}\beta_{nm}(g-g_c)^m.
\label{expbeta}
\end{eqnarray}
Then, plugging (\ref{expF}) and (\ref{expbeta}) into (\ref{nlRG2}), we obtain
\begin{eqnarray}
\beta_{10}=0, \quad \gamma=\frac{2}{\beta_{11}+4a_2 \beta_{20}},
\label{nlbg}
\end{eqnarray}
and
\begin{align}
2a_0=&\beta_{00},  \qquad
2a_1=\beta_{01}, \qquad
2a_2=\beta_{02}+\beta_{11}(2a_2)+\beta_{20}(2a_2)^2. 
\label{nla012}
\end{align}
First equation of (\ref{nlbg}) determines the $g_c$, and the second determines the critical exponent $\gamma$.
$a_1$ and $a_2$ are determined by (\ref{nla012}).

Let us calculate the critical exponent by these formulas with (\ref{nlbetas}).
First, the fixed point $g_c$ is determined by $\beta_{10}=0$ equation.
It, however, involves $a_1$.
We recall that the non-linear contribution  $b_2$ in $\beta_{1}$ is
fairly small for the perturbative regime,
because it is proportional to $g^5$ and hence negligible with a good accuracy
when we determine the fixed point.
Thus, we take $g_c\simeq -0.113$ that is determined by the linearized RG equation
as the fixed point of $\beta_{10}$. 
With $g_c\simeq -0.113$, from the second equation of \eqref{nla012},
one finds $a_1=2.49$ and $55.6$. We use them to compute $a_2$ and $\gamma$.
Under the weak coupling limit $g \rightarrow 0$,
$b_2'$ tends to be zero, and it is found that
the smaller branch of $a_1$ has a smooth limit. 
If $a_1$ is not singular, then   
$\beta_{20} \rightarrow 0$ and $\beta_{10}\rightarrow 0$ while
$\beta_{11}$ and  $\beta_{02}$ stay finite under the limit.
Therefore, from the  \eqref{nla012},
one can see that the smaller branches of $a_1$ and $a_2$ have smooth $g \rightarrow 0$ limits.
The corresponding $a_2$ value is calculated to be $69.1$, and
the exponent is found to be $\gamma=1.59$.
We, thus, adopt $\gamma=1.59$ for $a_1=2.49$ and $a_2=69.1$
 as the value of the critical exponent through the nonlinear analysis.


\section{A solution of (\ref{yypsd})}
\label{sec:an-exact-solution}

We are going to solve (\ref{yypsd}) with the condition $A=B=0$.
In this case, the tree level propagator is diagonal,
$\Upsilon^{-1}_{AB}= \Delta_{AB}{\bf 1}$ with
\begin{align}
\Delta_{aa} =& (1+\lambda\alpha^2) \Delta^{-1}
\,, \quad
\Delta_{bb} = (1+\lambda\beta^2) \Delta^{-1} \,,
\quad
\Delta_{ab}=\Delta_{ba} =
(\lambda\alpha\beta) \Delta^{-1},
\nn\\
& \Delta =
(1+\lambda\beta^2)(1+\lambda\alpha^2)-\lambda^2\alpha^2\beta^2 \,.
\end{align}
We now assume that $C_{AB}=c_{AB}{\bf 1}$ and solve the equations.
For further simplification,
we set $\beta=0$ and keep $\alpha$ to be non-zero (or vice versa),
and now
 $\Delta_{ab}=\Delta_{ab}=0$.
(\ref{yypsd}) is reduced to
\begin{eqnarray}
c_{aa}&=& \Delta_{aa}-\lambda\Delta_{aa}c_{aa}c_{bb}
-\lambda\Delta_{aa}\left( c_{ba}- 2 c_{ab} \right)c_{ba}, 
\nonumber \\
c_{bb}&=& \Delta_{bb}-\lambda \Delta_{bb}c_{aa}c_{bb} 
-\lambda\Delta_{bb}\left( c_{ab}-2 c_{ba} \right)c_{ab}, 
\nonumber \\
c_{ab} &=& \lambda\Delta_{aa}c_{bb} \left(c_{ab} - c_{ba} \right), 
\nonumber \\
c_{ba} &=& - \lambda\Delta_{bb}c_{aa} \left(c_{ab} - c_{ba} \right).
\label{yypsdap}
\end{eqnarray}
From the last two equations in (\ref{yypsdap}),
in order for a solution with non-zero $c_{ab}$ and $c_{ba}$ to exist,
\begin{eqnarray}
\lambda \left( \Delta_{aa}c_{bb} + \Delta_{bb}c_{aa} \right) =1 \,.
\end{eqnarray}
In the weak coupling limit $\lambda\rightarrow 0$, 
the exact propagators are to be the tree level ones, $c_{aa}=c_{bb}=1$,
and this condition cannot be satisfied.
Therefore, in order to have solutions that have smooth $\lambda\rightarrow 0$ limit,
$c_{ab}=c_{ba}=0$.
Since the tree level propagator does not have off-diagonal components,
this is consistent with the weak coupling limit.
Now the solutions for the rest are  easily found as
\begin{eqnarray}
c_{aa}&=&\frac{1+\lambda\alpha^2}{2\lambda}(-1+\sqrt{1+4\lambda(1+\lambda\alpha^2)^{-1}}), \nonumber \\
c_{bb}&=&\frac{1}{2\lambda}(-1+\sqrt{1+4\lambda(1+\lambda\alpha^2)^{-1} }) \,.
\end{eqnarray}

\section{Schwinger-Dyson equations of Yang-Mills matrix model}
\label{sec:schw-dyson-equat}
We study Schwinger-Dyson equations of the
generalized model (\ref{Ymmm2}) with $p=q=0$:
\begin{eqnarray}
S = \tr\left(\frac{m}{2}A^2+\frac{M}{2}B^2+gA^2B^2+hABAB\right).
\label{eq:8}
\end{eqnarray} 
Note that when $h=-g$, this model comes back to the original
Yang-Mills matrix model with the mass terms \eqref{YMmm}.
In this appendix, we do not put the ``hat'' symbol on the matrices. 
Under the reparametrization $A \rightarrow A+\delta A$ and
$B \rightarrow B+\delta B$, the invariance of the partition function
leads to the following Schwinger-Dyson equations,
\begin{align}
  \label{eq:7}
  \int DADB \, \tr \left[
\frac{\partial}{\partial A} \left(
\delta A e^{-NS} \right) \right] =0 \,, 
\qquad
  \int DADB \, \tr \left[
\frac{\partial}{\partial B} \left(
\delta B e^{-NS} \right) \right] =0 \,.
\end{align}
With $\delta A=A^3$, $\delta B=A^2 B$, and $\delta B=ABA$, 
we obtain
\begin{align}
2\vev{\tr A^2}+
\frac{1}{N}\vev{(\tr A)^2}
&=\vev{\tr \left(  mA^4+2gA^4B^2+2hA^3BAB \right)}, \nn \\
\vev{\tr A^2} =&
\vev{\tr \left( MA^2B^2+gA^4B^2+gA^2BA^2B+2hA^3BAB
\right)}, \nn \\
\frac{1}{N}
\vev{(\tr A)^2} =&
\vev{\tr \left( MABAB+2gA^3BAB+2hA^2BA^2B \right)}. 
\label{yysd2}
\end{align}
In this expression, again $\vev{(\tr A)^2}$ is being subleading due to
the large-$N$ factorization property and $\vev{ \tr A}=0$ with respect
to the action \eqref{eq:8}, and is therefore dropped.
When $g=\pm h$, one can eliminate all the terms involving six
matrices,
and obtain
\begin{eqnarray}
m\vev{\tr A^4}=2M \vev{\tr A^2B^2 }\mp M \vev{\tr ABAB} \,,
\qquad (h=\pm g)
\end{eqnarray}
Thus one of the quartic order operators, for example $A^4$, can be
written by the others. 
By considering the case with $\delta B=B^3$, $\delta A=B^2 A$ and $\delta A=BAB$, 
one finds the similar relation with
 $A \leftrightarrow B$ as well as $m \leftrightarrow M$,
\begin{align}
  \label{eq:9}
  M\vev{\tr B^4}=2m \vev{\tr A^2B^2 }\mp m \vev{\tr ABAB} \,.
\qquad (h=\pm g)
\end{align}
In the main part, we use the these relations with $h=-g$.

\section{A matrix analogue of the mean field approximation}     
\label{sec:matrix-analogue-mean}

The mean field approximation is a useful tool
in the statistical mechanics.
Here, we discuss a matrix model analogue of it.
In this appendix, we omit the ``hat'' symbol on the matrices.

First, we consider the $\phi^4$ model and
 introduce a ``mean field''  $\sigma$ as
\begin{eqnarray}
\sigma=\vev{ \frac{1}{N}\tr\phi^2} \,.
\end{eqnarray}
We assume that the mean field $\sigma$ takes a non-zero value
and replace the $\tr \phi^4$ term in the original action (\ref{p4action})
with $k \sigma \; \tr \phi^2$, 
where $k$ is a constant number determined by combinatrics.
A natural choice would be $k=6$, namely, $2$ out of $4$,
but here we take $k=4$.
This is the number of the adjacent pairs in the trace,
and may be interpreted as a ``planar'' paring.
Under this replacement, the action and the partition
function read
\begin{align}
 S \rightarrow &
\frac{1}{2} \left(1+2g\sigma  \right) \tr \phi^2 \,,
\qquad
Z[\sigma] \propto (1+2g\sigma)^{-N^2/2} \,.
\end{align}
The mean field should satisfy the consistency condition:
\begin{align}
\sigma =&
\frac{1}{N} \vev{\tr \phi^2} = -\frac{1}{N^2} \frac{1}{g} \frac{\partial}{\partial \sigma}
\log Z[\sigma] = \frac{1}{1+2g\sigma} \,.
\end{align}
This is solved as
\begin{eqnarray}
\sigma=\frac{1}{4g}(-1+\sqrt{1+8 g})\,,
\label{meanp4}
\end{eqnarray} 
where the sign is chosen so that $\sigma$ stays finite as $g \rightarrow 0$. 
It develops the singularity around $g_c = -1/8$, which behaves as $(g-g_c)^{1/2}$.
We compare (\ref{meanp4}) with the exact solution
by Brezin et.~al. \cite{Brezin:1977sv},
\begin{eqnarray}
\frac{1}{N}\vev{ \tr\phi^2 }
=\frac{(-1+\sqrt{1+12g})^2(1+2\sqrt{1+12g})}{108g^2}.
\label{exactp2}
\end{eqnarray}
One can see that singularity structure is $(g-g_c^\text{exact})^{3/2}$
with $g_c^\text{exact}=-1/12$.
The mean field approximation captures the critical behavior
qualitatively.
It also reproduces reasonable approximated values of the critical coupling constant
and the order of the singularity.

Next, we move on to the Yang-Mills matrix model with the mass terms \eqref{YMmm}.
We may start with a generalized one, (\ref{Ymmm2}), which comes back to
\eqref{YMmm} by setting $p=q=0$ and $h=-g$.
Mean fields are introduced as\footnote{%
$\sigma_{ab}=0$ with respect to the action \eqref{Ymmm2}.
We introduce it to explore the possibility of
the $Z_2$ symmetry, $A\rightarrow -A$ or  $B\rightarrow -B$,
being spontaneously broken.}
\begin{eqnarray}
\sigma_{a}=\frac{\langle \tr A^2 \rangle}{N}, \quad 
\sigma_{b}=\frac{\langle \tr B^2 \rangle}{N}, \quad 
\sigma_{ab}=\frac{\langle \tr AB \rangle}{N}. 
\end{eqnarray}
We replace the quartic order terms in (\ref{Ymmm2}) as
\begin{align}
\tr A^2B^2 \rightarrow &
\sigma_a \tr B^2+\sigma_b \tr A^2+2\sigma_{ab} \tr AB, \quad
\tr ABAB \rightarrow 4 \sigma_{ab} \tr AB, \nn\\
\tr A^4 \rightarrow & 4\sigma_{a} \tr A^2, \quad 
\tr B^4\rightarrow 4\sigma_{b} \tr B^2 \,,
\nn
\end{align}
where we have again assumed ``planar'' parings.
The action (\ref{Ymmm2}) and the partition function
are
\begin{align}
&
S\rightarrow 
\frac{1}{2} (m+2g\sigma_b+2p\sigma_a) \tr A^2
+\frac{1}{2} (M+2g\sigma_a+2q \sigma_b) \tr B^2
+(2g+4h) \sigma_{ab} \tr AB  \,,
\\ &
Z[\sigma_a,\sigma_b,\sigma_{ab}] \propto 
\left\{
(m+2g\sigma_b+2p \sigma_a)(M+2g\sigma_a+2 q\sigma_b)-4 (g+2h)^2 \sigma_{ab}^2
\right\}^{-\frac{N^2}{2}} \,.
\end{align}
The self-consistency equations read
\begin{align}
\sigma_{a} =& -\frac{2}{N^2} \frac{\partial}{ \partial m  }\log Z, 
\quad
\sigma_{b} = -\frac{2}{N^2} \frac{\partial}{ \partial M  }\log Z, \quad
\sigma_{ab} = -\frac{1}{N^2}\frac{1}{2g+4h}
\frac{\partial}{ \partial \sigma_{ab}  }\log Z.
\end{align}
For the Yang-Mills matrix model in question,
$p=q=0$ and $h=-g$, the equations are
\begin{align}
  \label{eq:22}
  \sigma_a=& (M+2g\sigma_b) D^{-1} \,,
\quad
\sigma_b= (m+2g\sigma_b) D^{-1} \,,
\quad
\sigma_{ab}= 2 g \sigma_{ab} D^{-1} \,,
\nn\\&
D=(m+2g\sigma_b)(M+2g\sigma_a)-g^2 \sigma_{ab}^2 \,.
\end{align}
A solution with $\sigma_{ab} \neq0$ exists only when $m=-M$, in which we are not interested.
For $\sigma_{ab}=0$, $m\sigma_a=M\sigma_b=\sigma$ satisfies the same equation
\begin{eqnarray}
\sigma=\frac{1}{1+2\lambda\sigma},
\end{eqnarray}
where $\lambda=\frac{g}{mM}$.
Thus, $\sigma$ has an equivalent solution,
\begin{eqnarray}
\sigma=\frac{1}{4\lambda}(-1+\sqrt{1+8\lambda}),
\end{eqnarray}
to the $\phi^4$ model.
As the mean-field approximation works qualitatively well in the $\phi^4$ case,
this result suggests that the Yang-Mills matrix model with the mass term \eqref{YMmm}
exhibits a similar critical behavior for small negative $\lambda$.

We may comment on a generalized model \eqref{Ymmm2}.
If $p$ and $q$ are non-zero,
$\sigma_a$ and $\sigma_b$ solve two third order equations
and will develop
different type singularities.


\end{document}